\documentclass[useAMS,usenatbib]{mn2e}
\usepackage[dvips]{graphicx}

\title[Differential Evolution of UVLF of LBGs]{Differential Evolution of the UV
Luminosity Function of Lyman Break Galaxies from $z \sim$ 5 to 3
\thanks{Based on data collected at Subaru Telescope and partly obtained from the SMOKA science archive at Astronomical Data Analysis Center, which are operated by the National Astronomical Observatory of Japan.}}
\author[I. Iwata et al.]{I. ~Iwata$^{1,2}$\thanks{E-mail:
iwata@oao.nao.ac.jp (II)},
K. ~Ohta$^2$,  N. ~Tamura$^{3,4}$,
M. ~Akiyama$^3$, K. ~Aoki$^3$, 
M. ~Ando$^2$, \and G. Kiuchi$^2$ and M. Sawicki$^{2,5,6}$\\
$^1$Okayama Astrophysical Observatory, National Astronomical
Observatory of Japan, Kamogata, Asakuchi, Okayama, Japan 719-0232\\
$^2$Department of Astronomy, Graduate School of Science, Kyoto University, 
Sakyo-ku, Kyoto, Japan 606-8502\\
$^3$Subaru Telescope, National Astronomical Observatory of Japan, 
650 North A`ohoku Place, Hilo, HI 96720\\
$^4$Department of Physics, University of Durham, 
South Road, Durham DH1 3LE, UK\\
$^5$Physics Department, University of California, Santa Barbara, 
CA 93106, USA\\
$^6$Department of Astronomy and Physics, St. Mary's University, 
923 Robie St., Halifax, Nova Scotia, B3H 3C3 Canada
}

\begin{document}

\date{Accepted 2007 January 29. Received 2007 January 22; in original form 2006 November 14}

\pagerange{\pageref{firstpage}--\pageref{lastpage}} \pubyear{2006}

\maketitle

\label{firstpage}

\begin{abstract}
We report the UV luminosity function (LF) of Lyman break galaxies at
$z \sim 5$ derived from a deep and wide survey using the prime focus
camera of the 8.2m Subaru telescope (Suprime-Cam).  Target fields
consist of two blank regions of the sky, namely, the region including
the Hubble Deep Field-North and the J0053+1234 region, and the total
effective surveyed area is 1290 arcmin$^2$.  Applications of carefully
determined colour selection criteria in $V-I_c$ and $I_c-z'$ yield a
detection of 853 $z \sim 5$ candidates with $z'_\mathrm{AB} < 26.5$
mag.  The UVLF at $z \sim 5$ based on this sample shows no significant
change in the number density of bright ($L \ga L^\ast_\mathrm{z=3}$)
LBGs from that at $z \sim 3$, while there is a significant decline in
the LF's faint end with increasing lookback time.  This result means
that the evolution of the number densities is differential with UV
luminosity: the number density of UV luminous objects remains almost
constant from $z\sim5$ to 3 (the cosmic age is about 1.2 Gyr to 2.1
Gyr) while the number density of fainter objects gradually increases
with cosmic time.  This trend becomes apparent thanks to the small
uncertainties in number densities both in the bright and faint parts
of LFs at different epochs that are made possible by the deep and wide
surveys we use.
We discuss the origins of this differential evolution of the UVLF
along the cosmic time and suggest that our observational findings are
consistent with the biased galaxy evolution scenario: a galaxy
population hosted by massive dark haloes starts active star formation
preferentially at early cosmic time,
while less massive galaxies increase their number density later.  We
also calculated the UV luminosity density by integrating the UVLF and
at $z \sim 5$ found it to be
$38.8^{+6.7}_{-4.1}$\% of that at $z \sim 3$ for the luminosity range
$L>0.1 L^\ast_\mathrm{z=3}$.  By combining our results with those from
the literature, we find that the cosmic UV luminosity density marks
its peak at $z=2$--3 and then slowly declines toward higher redshift.
\end{abstract}

\begin{keywords}
cosmology: observations -- galaxies: high-redshift --
galaxies: luminosity function -- galaxies: evolution -- galaxies: formation.
\end{keywords}

\section{Introduction}

The galaxy luminosity function (LF) is one of the fundamental
quantities frequently used to investigate the properties of galaxy
populations and explore the evolutionary process of galaxies.  It
represents the number densities of objects with specific luminosity at
specific redshift ranges.  Recent development of large imaging and
spectroscopic surveys improved the accuracy of LFs in the local
universe in optical and near-infrared wavelengths (\citealt{2dflf};
\citealt{sdsslf}; \citealt{2masslf}).  Good quality LFs at UV wavelengths 
have also been obtained by combinations of balloon or spaceborne
missions with ground-based survey data (\citealt{sul00};
\citealt{wyd05}).  For LFs at earlier cosmic times, in principle a complete and
volume-limited sample of galaxies at the target redshift is required.
Great efforts have been made by spectroscopic surveys to obtain large
galaxy samples capable of deriving LFs up to $z \la 2$ (e.g.,
\citealt{cfrslf}; \citealt{cnoc2lf}; \citealt{vvdslf1}). 
However, a construction of a large spectroscopic sample of galaxies at
higher redshift, $z >4$, would require enormous observing time due to
the faintness of the target objects and, consequently, it remains a
big issue to be addressed by future observing facilities. Instead of
determining redshifts of large numbers of objects, we are able to
derive a luminosity function statistically by using a sample of
galaxies with redshift estimates based on their spectral energy
distributions traced by multi-band photometry.  In this way, extensive
multi-wavelength surveys conducted by the Hubble Space Telescope (HST)
and large ground-based telescopes have enabled us to measure LFs at
different epochs up to $z \la 2$--3 (e.g.,
\citealt{sly97}; \citealt{dro03}; \citealt{sdf3}; \citealt{chen03}; 
\citealt{poz03}; \citealt{wol03}; \citealt{dah05}) 
and earlier (e.g., \citealt{lanz02}; \citealt{poli03};
\citealt{gab04a}; \citealt{gab06}; \citealt{pal06}), 
and to discuss the evolution of galaxies through the comparison of LFs
at different cosmic times.

One of the most powerful observational methods to probe galaxies at
high redshift ($z \ga 2$) is the detection of the spectral
discontinuity due to the redshifted Lyman limit (912$\mathrm \AA$) or
Lyman $\alpha$ (caused by absorption by inter-galactic hydrogen gas)
through multi-wavelength broad-band imaging.  This method is called
the Lyman break method, and since the pioneering work by Steidel and
his co-workers (\citealt{S92}; \citealt{S96}), large number of Lyman
break galaxies (LBGs) at $z \approx 2$--4 have been detected (e.g.,
\citealt{mad96}; \citealt{steidel03}; \citealt{steidel04}; 
\citealt{fou03}; \citealt{kdf1}).

There are several obstacles in searching for higher redshift ($z>4$)
LBGs.  First, the larger distance makes the apparent brightness of
target objects dimmer so that, for the same luminosity, an object at
$z=5$ is about 1 magnitude fainter than that at $z=3$.  Next, the
sensitivity of CCD chips declines at $\lambda > 8000
\mathrm{\AA}$.  And finally, for ground-based instruments, the
increase of background sky brightness due to the night sky lines
emitted from the Earth's atmosphere makes the S/N much worse than in
the bluer wavelength ranges.  Despite these difficulties, it is
important to push to these earlier epochs in the history of galaxy
formation.

In Iwata et al. (2003; hereafter \citealt{i03}) we created a sample of
$\approx 300$ LBGs with $I_c\mathrm{(AB)}<26.0$.  It was the first
large sample of LBGs at $z \sim 5$, and was made possible by the
unique combination of the large mirror aperture of the 8.2m Subaru
telescope and the wide field of view of its prime focus camera named
Suprime-Cam \citep{miy02}.  It was also in this work that the UVLF of
LBGs at $z \sim 5$ was derived statistically for the first time.  We
found no significant change in the bright luminosity range
($M_\mathrm{UV} \la -21$ mag) between $z=5$ and $z=3$.  After
\citet{i03} several studies have successfully created LBG samples at
$z \sim 5$ (\citealt{leh03}; \citealt{sdf5};
\citealt{gia04b}; \citealt{lee06}). Attempts to obtain 
statistically meaningful samples of LBGs at $z \ga 6$ have also been
fruitful (e.g., \citealt{bou03}; \citealt{bou06a}; \citealt{bou06b};
\citealt{sta03}; \citealt{dic04}; \citealt{bun06}).
As we accumulate knowledge of the physical properties of high-redshift
($z > 2$) galaxies such their as sizes (e.g., \citealt{S96};
\citealt{gia96}; \citealt{rav06}), stellar populations (e.g., 
\citealt{sy98}; \citealt{shap01}; \citealt{pap01}; \citealt{eyl05}; 
\citealt{yan06})  and clustering properties 
(e.g., \citealt{gia98}; \citealt{ade98}; \citealt{sdf6};
\citealt{ham04}; \citealt{kas06}), we improve our chances of 
constructing a comprehensive view of galaxy evolutionary processes in
the high redshift universe and eventually through cosmic time.

One serious limitation of past ground-based LBG surveys is that only
luminous objects have been studied. The statistical nature of galaxies
fainter than the typical luminosity ($L^\ast$ in the LF parameterized
with the \citealt{sch76} form) has been left unexplored, although these
numerous faint galaxies are estimated to dominate the rest-frame UV
luminosity density.  Thus, a comparison between the evolution of
luminous objects and that of faint objects has not been made.

The Keck Deep Fields (KDF) \citep{kdf1} made a very deep imaging
survey reaching ${\cal R}\mathrm{(AB)} \sim 27$ mag and constructed
samples of LBGs at $z\sim$2 to 4 including objects far fainter than
$L^\ast$.  They find that the UVLF at $z \sim 4$ shows a significantly
smaller number density in the fainter part than that at $z \sim 3$,
while in the bright end no difference between $z\sim4$ and 3 has been
found.  Such differential evolution of the UVLF was not detected in
the preceding LBG work with brighter limiting magnitudes
\citep{steidel99}. The KDF indicated that deep surveys reaching
fainter than $L^\ast$ (with sufficient survey area) are required to
closely inspect the nature of high-z galaxy evolution.  It is an
interesting subject to examine whether such differential evolution
started at earlier epoch, $z \ga 5$.  For that subject the
construction of deep sample of LBGs down to 1 magnitude fainter than
$L^\ast$ at $z \ga 5$ is indispensable.

Here we report the UVLF of LBGs at $z \sim 5$ based on our updated
sample that builds on the work of I03.  The survey now consists of two
independent fields and has an effective survey area of 1290 arcmin$^2$
after masking bright objects. The limiting magnitudes of the sample
LBGs are $z'=26.5$ mag for one field (the field including the Hubble 
Deep Field - North) and 25.5 mag for the other (J0053+1234).  
The sample limiting
magnitude is now $\sim 0.5$ mag deeper than in our previous work
(I03), which was 26.0 mag in $I_c$.  The survey area is more than two
times larger than our previous survey, which enables us to examine the
degree of field-to-field variance and to estimate how representative
are our measured properties of $z \sim 5$ galaxies.  Through the
comparison of the $z \sim 5$ UVLF obtained with this deeper and wider
sample with the deep UVLFs at $z \sim 3$ and 4 by \citet{kdf2}, we
attempt to disclose the evolution in the UVLFs of LBGs.

In section 2 we describe the observations for the two target
fields. Procedures of data processing, catalog construction and sample
selection are described in section 3.  In section 4 we present our $z
\sim 5$ UVLFs and also compare them with previous results. 
In section 5 we make a comparison of our $z \sim 5$ UVLF with those at
$z = 4$ and 3 in \citet{kdf2} and then discuss the differential
evolution of UVLF.  Then in section 6 we derive the UV luminosity
density at $z \sim 5$ and discuss the evolution of the UV luminosity
density through the cosmic time. Our findings are summarized in
section 7.  Throughout this paper we adopt a cosmology with the
parameters $\Omega_M=0.3$, $\Omega_\Lambda=0.7$, $H_0=70$ km/s/Mpc.
Magnitudes are on the AB system.

\section{Observations}

We chose two target fields for our survey of LBGs at $z \sim 5$.  One
is in the area of the Hubble Deep Field - North (HDF-N;
\citealt{wil96}), and the other is J0053+1234 (\citealt{coh96};
\citealt{coh99a}).  Both fields are target fields of existing
extensive redshift surveys (e.g., \citealt{coh00}; \citealt{daw01};
\citealt{tk04}; \citealt{cow04}; \citealt{coh99b}), 
and there are many galaxies with spectroscopic identifications at
intermediate redshifts ($z<4$).  Such information is valuable to
define reliable colour selection criteria for LBGs at our target
redshift range, $z=5$.

\begin{figure}
\resizebox{8cm}{!}{
\includegraphics{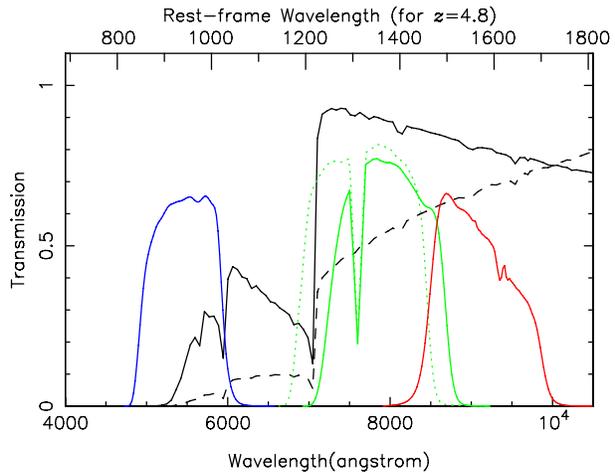}}
\caption{
Effective transmissions (including quantum efficiency of the CCD and
atmospheric absorption) for the three filters used in our observations
(solid lines with blue ($V$), green ($I_c$) and red($z'$)). The
transmission of $i'$ filter used in \citet{sdf5} is also shown as a
dotted line.  Model spectra of star-forming galaxies at $z=4.8$ are
shown as black lines. The solid line represents the case without dust
attenuation and the dashed line is for the case with
$E(B-V)=0.4$. These SEDs are normalized at 10,000 \AA. See text for
details of the models. Upper-side ticks indicate rest-frame
wavelengths for an object at $z=4.8$.  }
\label{fig:filters}
\end{figure}

We use the $V$, $I_c$ and $z'$ filters to select our $z\sim 5$ LBGs.
In Fig.~\ref{fig:filters} we show the transmissions of these filters
as a function of wavelength.  In the figure we also indicate model
SEDs of star-forming galaxies with and without dust attenuation that
we generated using version 2 of the population synthesis code PEGASE
\citep{fio97}.  A constant star formation rate with the Salpeter
IMF \citep{sal55} is assumed, and the age from the onset of star
formation is 100 Myr.  Dust attenuation we show here is $E(B-V)=0.4$
using the attenuation law proposed by \citet{cal00} and absorption by
intergalactic neutral hydrogen is applied using the analytic formula
in \citet{ino05}.  The redshift we assumed here is 4.8, which is an
average redshift of our sample LBGs estimated with simulations (see
section 4.1). In this figure the model SEDs are normalized at 10,000
\AA\ in the observer frame.  The $V I_c z'$ filter combination 
efficiently detects the break at Ly$\alpha$ for star-forming galaxies
at $4.3 \la z \la 5.3$, as verified by simulations (described in
section 4.1) and follow-up spectroscopy (\citealt{and04}; 
\citealt{and07}).  In addition to
our $V I_c z'$ filters, Fig.~\ref{fig:filters} also shows the
$i'$-band filter (dotted line) and we note that \citet{sdf5} use the
combination of $V$, $i'$ and $z'$ filters to select their LBGs at
$z\sim 5$.  The $i'$-band filter is shifted slightly
shortward compared to the $I_c$-band filter used in our observations.

\subsection{The Hubble Deep Field - North}

We made deep imaging observations of the field that surrounds the
HDF-N using the Suprime-Cam attached to the 8.2m Subaru Telescope in
February 2001. The results presented in \citet{i03} are based on these
data alone.  For the new results we show in the present paper we also
used the imaging data taken by the University of Hawaii (UH) group
\citep{cap04} using the same instrument and the same filters during
February 28 -- April 23, 2001.  We retrieved these UH data from the
SMOKA data archive \citep{bab02}.

Properties of these two sets of observations are different from each
other in several aspects.  First, the central field positions are
$\sim$3\arcmin~ apart, and in our observations the position angle was
set to 10\degr (east of north), while in the observations by the UH
group position angles were 0\degr or 180\degr. The position and the
position angle in our observation were adjusted so that the HDF-N
taken by the HST/WFPC2 was located at the centre of one of
Suprime-Cam's CCD chips.  The dithering lengths are also different. In
our observations dithering scales are only 10\arcsec--15\arcsec, which
were chosen so that the area with deepest integration time became as
large as possible. We could choose such small dither lengths because
we concentrated on the detection of galaxies at $z \sim 5$, which
should be apparently small ($\la 2\arcsec$).  In the UH observations
by \cite{cap04} the dither scales were around 1\arcmin and the
effective exposure time within the observing field is fairly close to
uniform.  In this paper the area common to these two different sets of
observations is used, since that area achieves the deep image depths
in all of three observing bands ($V$, $I_c$ and $z'$).  In
Table~\ref{tbl:hdfnobs} we summarize observations for the HDF-N
region.

\subsection{J0053+1234}

In order to increase the survey area and examine the effect of
field-to-field variance, we carried out the imaging of another blank
field.  Specifically, we imaged the J0053+1234 region, which is one of
the target fields of the Caltech Faint Galaxy Redshift Survey
(\citealt{coh99a}; \citealt{coh99b}).  Using extensive Keck
spectroscopy, these authors identified 163 galaxies in the redshift
range $0.173 < z < 1.44$ within an $2\arcmin \times 7.\arcmin3$ field.
Steidel et al. (1999, 2003)
also selected this region as one of the 
target fields of their searches for LBGs at $z \sim 3$ and 4, 
and they identified 19 galaxies with spectroscopic redshifts 
at $3 < z < 5$.

Our imaging observations of this field using Subaru / Suprime-Cam have
been executed during the summer seasons from 2002 to 2004.  The
central coordinate of the pointing is 00$^h$53$^m$23.0$^s$,
+12\degr33\arcmin56\arcsec (J2000.0).  In Table~\ref{tbl:j0053obs} we
summarize information of our observations and final limiting
magnitudes in $V$, $I_c$ and $z'$.  Although the point spread function
of the final images are smaller than that in our images of the HDF-N
region, the total limiting magnitudes achieved for this field are 0.3
-- 0.5 mag shallower than those for the HDF-N region since the
effective net exposure times are shorter due to the poor weather
conditions we experienced. Nevertheless, these data more than double
the total survey area and so are very valuable for populating the
relatively rare, bright part of the sample and for examining
field-to-field variations.

\section{Data Reduction and Selection of LBG Candidates}

\subsection{Image Processing}

The basic image processing was done using nekosoft --- the software
developed for the data reduction of mosaiced CCD cameras
\citep{yagi02} --- as well as with  IRAF. Bias was subtracted 
using overscan regions.  Next, flat-field frames were generated by
combining median-normalized object frames after masking objects. Then,
after division by the flat-field frame and cosmic-ray removal (using
CRUTIL in IRAF), a distortion correction was made for every frame with
correction factors which were determined for each CCD position.

Image mosaicing in each band was made by identifying 40--180 stars
common in object frames.  The correction for differences in
sensitivity and fluctuations of point source flux was also made based
on these star data.  In the HDF-N region, the data taken by us have
small dither scales and so we could not adopt a mosaicing procedure
usually executed in nekosoft, which determines relative positions of
frames by common stars in different CCD chips.  Thus, we first derived
relative positions of frames taken by the UH group and generated a
mosaiced image using the UH data alone.  We then calculated the
positions of the frames taken by us relative to that mosaiced
image. The final mosaiced image was then made using both the UH and
our data. Among the 10 CCD chips of the Suprime-Cam, one chip was not
working during part of the observations of the HDF-N region.  We also
excluded data of another CCD chip in our data because there is not
enough overlapping area between the positions of this chip and the UH
data to allow us to determine reliable relative positions.
In the J0053+1234 region data from all CCD chips were used.
The effective survey areas (after masking bright objects) are 508.5
arcmin$^2$ (the HDF-N region) and 781.4 arcmin$^2$ (the J0053+1234
region).  The FWHM values of point sources in the mosaiced images are
about 1.1\arcsec for the HDF-N region and 0.9\arcsec for the
J0053+1234 region (Tables~\ref{tbl:hdfnobs} and \ref{tbl:j0053obs}).

\subsection{Astrometry}

The conversions from pixel coordinates in mosaiced CCD images 
to equatorial coordinates were made using the USNO-B1 catalog 
\citep{mon03}. In total 1,000 to 1,300 stars in the catalog were 
identified in the mosaiced image in each band, and polynomial
coefficients up to third order were calculated.  Internal position
errors in coordinate conversions are less than 0.2\arcsec rms all over
the images. Because these accuracies are almost identical to that of
the USNO-B1 catalog itself, we expect that positional uncertainty due
to the errors in the corrections for optical image distortions and
those in the mosaicing process to be smaller than 0.2\arcsec.

\subsection{Photometric Catalog}

The determination of photometric zero points for the HDF-N region is
described in \citet{i03}.  Briefly, we determined photometric zero
points based on imaging data of \citet{land92} standard stars for $V$
and $I_c$ bands, while for $z'$-band the zero point was determined
through the distribution of $I_c-z'$ colours of field stars and
galaxies exposed on our images.  Systematic errors in the
determination of zero points are estimated to be 0.02, 0.08 and 0.1
mag for $V$, $I_c$ and $z'$-bands, respectively.  The photometric zero
points of the final mosaiced images were determined by comparison with
the images used in \citet{i03}.  For the J0053+1234 region, the same
procedure was followed for the $V$ and $I_c$ bands, while for
$z'$-band images of spectrophotometric standard stars taken during the
same observing nights were used.  Systematic errors are estimated to
be 0.04 mag for $V$-band and 0.03 mag for $I_c$ and $z'$-bands.
Corrections for Galactic extinction were made based on the dust map by
\citet{sch98}. The amount of reddening for the HDF-N region is assumed to
be $E(B-V)=0.012$, and for the J0053+1234 region it is $E(B-V)=0.066$.

Object detection and photometry were made by using SExtractor
\citep{ber96} ver. 2.3 in single-image mode.  A Gaussian filter 
with FWHM=3 pixels was used, and at least 4 contiguous pixels with
counts higher than 1.3 $\sigma$ above background counts were required
to be detected as an object.  The cross-identifications of objects in
our three bands were made in equatorial coordinates.  Objects were
included in the object catalog only when they were within the area
with sufficient image depth, determined by checking the exposure maps
which take account of the number of exposures and variations in
sensitivity between individual CCD chips.  This procedure guarantees
that the image quality is fairly close to uniform and that the
variance of photometric errors does not affect the uniformity of the
galaxy sample.  For $V$-band dropout LBGs we require objects to be
detected both in $I_c$ and $z'$-bands.  We used \verb+MAG_AUTO+ as a
total magnitude in all bands, and derived $V-I_c$ and $I_c - z'$
colours using 1.6\arcsec diameter aperture magnitudes.  Since object
centroids are determined separately in each band, there is a
possibility that the positions of apertures in different bands have
offsets (e.g., star-forming regions are prominent in $I_c$-band and
their distribution is irregular).  We feel that such problems are
rare, since FWHM sizes of LBG candidates are mostly smaller than the
aperture size used to derive colours (1.6\arcsec).  We also find
through visual inspection that morphologies of $V$-dropout objects in
$I_c$-band and in $z'$-band are similar in most cases.  The
photometric catalogs of both fields were constructed based on the
$z'$-band.  For objects undetected by SExtractor in the $V$-band,
$V$-band aperture photometry was made at the objects' positions in the
$z'$-band. In these cases, if the flux density within 1.6\arcsec
diameter exceeded the 3-$\sigma$ sky fluctuation then the measured
aperture magnitude was used to calculate $V-I_c$ colour.  Otherwise
the 3-$\sigma$ upper limit was adopted.

\subsection{Selection of LBG Candidates}

The colour criteria we adopted to select LBG candidates at $z \sim 5$
are the same as those used in \citealt{i03}.  They were defined to
efficiently select star-forming galaxies at $z \sim 5$ without heavy
contamination by objects from lower redshifts, and were chosen by
considering the colours of both model galaxies and real galaxies
within our images which have spectroscopic redshifts identified by
previous surveys available at that time (see below for the
cross-identifications of galaxies with spectroscopic redshifts with
our object catalogues).  Our colour criteria are expressed by the
following equations in the AB magnitude system:

\[
 V-I_c > 1.55
\] and 
\[
 V-I_c > 7.0 (I_c - z') + 0.15.
\]

\begin{figure}
\resizebox{8cm}{!}{
\includegraphics{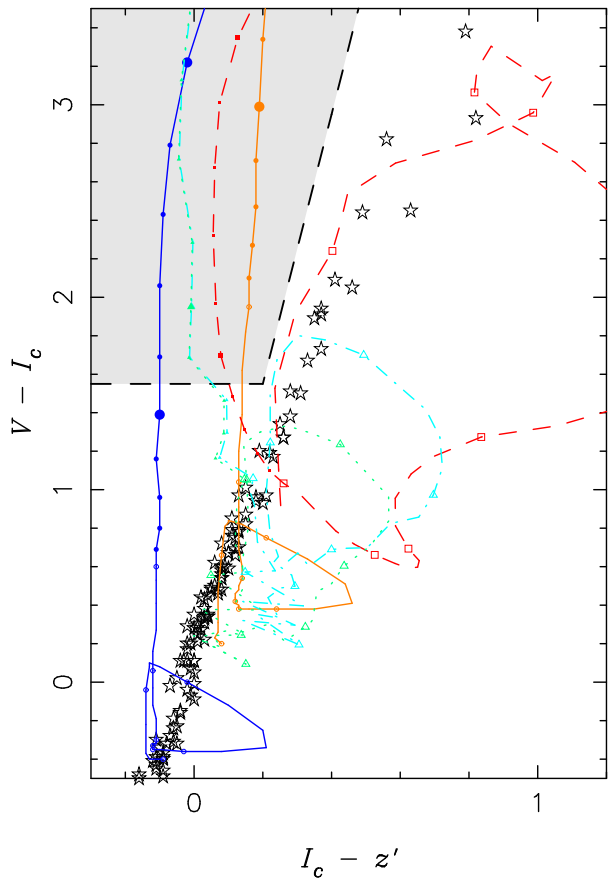}}
\caption{
A $V-I_c$ and $I_c-z'$ two colour diagram. 
The shaded area stands for the colour selection 
criteria adopted for the selection of Lyman 
break galaxies at $z\sim 5$.
Lines are loci of colour changes for model galaxies 
along redshift. Solid blue: a star-forming galaxy 
without dust attenuation. Solid orange: a star-forming 
galaxy with dust attenuation of $E(B-V)=0.4$. 
Solid circles are plotted for $z \ge 4.1$ with 
an interval of 0.1. Symbols at $z=4.5$ and $z=5.0$ 
are enlarged. 
Colour tracks calculated using template SEDs by 
\citet{cww80} are also shown. 
Cyan (dot-dashed): Sbc, light blue (dotted): Scd, 
red (dashed): elliptical. 
Symbols are plotted with a redshift interval of 0.5. 
Star symbols indicate the colours of A0--M9 stars calculated based on
the library by \citet{pick1998}. 
}
\label{fig:twocol0}
\end{figure}

In Fig.~\ref{fig:twocol0} we show our $z \sim 5$ colour selection
criteria along with redshift colour tracks of several types of
galaxies.  The colour tracks of model star-forming galaxies are
calculated based on the same SEDs as used in Fig.~\ref{fig:filters}
(see section 2).  Colour tracks for spiral (Sbc and Scd) and
elliptical galaxies are calculated using template SEDs of local
galaxies by \cite{cww80}.

\begin{figure*}
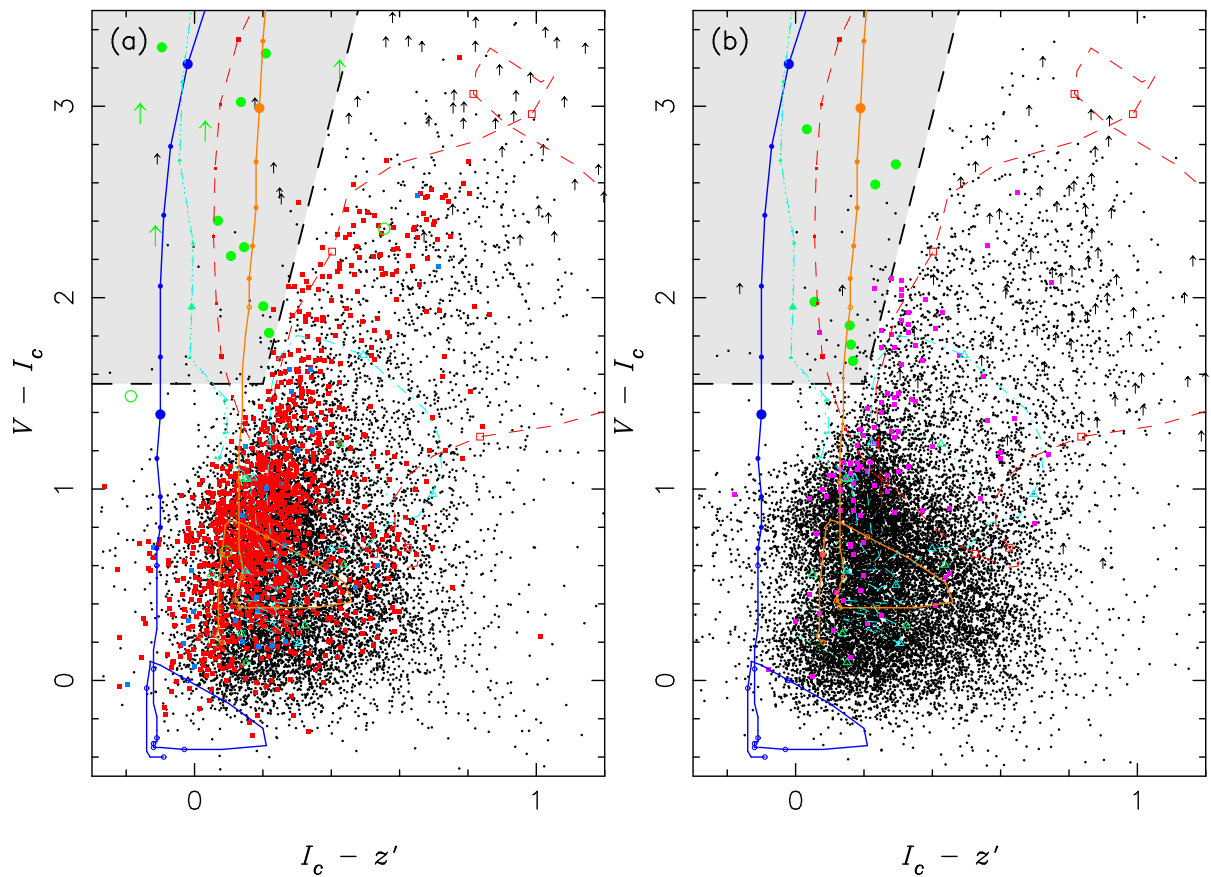

\resizebox{16cm}{!}{
\includegraphics{fig03a.eps}
\includegraphics{fig03b.eps}}
\caption{
$V-I_c$ and $I_c-z'$ two colour diagrams. 
(a): data points are for objects in the HDF-N region. 
Black circles indicate colours of objects 
with $23.0 < z' \leq 24.5$. For objects 
not detected in $V$-band, arrows are used to 
show the 3-$\sigma$ lower limits of $V-I_c$ colours. 
Red and blue squares represent galaxies with 
spectroscopic redshifts ($z<4.0$) published in 
\citet{tk04}
and \citet{coh00} 
respectively. Objects in \citet{coh00} 
which are listed in the catalog by \citet{tk04} 
are omitted. 
Colour track lines and star symbols are 
the same as in Fig.~\ref{fig:twocol0}.
Filled green circles indicate colours of galaxies at $z>4.5$ 
confirmed by spectroscopy. 
Three open green circles show colours of objects which 
are not included in our LBG candidates catalog but are reported 
to be at $z>4.5$ in previous spectroscopic observations 
(see table \ref{tbl:crossid_hdfn}). 
(b): same as (a) but for the J0053+1234 region.
Squares in magenta are objects with spectroscopic redshifts
smaller than 4.0 in the region, mostly taken from 
\citet{coh99a}, 
and green circles are galaxies at $z>4.2$ identified by 
spectroscopy by us and \citet{steidel99} 
(see table \ref{tbl:crossid_j0053}).
See text for more details of these figures.
}
\label{fig:twocol1}
\end{figure*}

In Fig.~\ref{fig:twocol1} we show colours of detected objects 
within a magnitude range $23.0 < z' < 24.5$.
The numbers of objects which satisfy the colour 
criteria are 617 and 236 for the HDF-N region ($z'<26.5$ mag) 
and the J0053+1234 region ($z'<25.5$ mag), respectively.

Among the HDF-N field LBG candidates in \citet{i03} that are within
the region common to the new, deeper and narrower imaging data, all of
the objects with $I_c \leq 24.5$ mag and 82\% of objects with $I_c
\leq 26.0$ mag are included in our  updated source catalog. 
The number of objects that were listed as LBG candidates in
\citet{i03} but  are outside the colour criteria in the new
catalog is 20--30\% larger than the estimated number of low-redshift
contaminants (see section 4.1 and Table~\ref{tbl:hdfuvlf}).  This fact
is not surprising, because the signal-to-noise ratios of our previous
data are worse than the HDF-N region data presented in this paper.

Both of our target fields have been observed with other intensive
spectroscopic surveys. We identified objects with spectroscopic
redshifts in our images and plotted their colours in
Fig.~\ref{fig:twocol1}. For the HDF-N region we used the results of
two surveys (\citealt{coh00} and \citealt{tk04}). The numbers of
objects identified are 644 for \citet{coh00} and 1,803 for
\citet{tk04} (most of the objects in \citealt{coh00} are included in
\citealt{tk04}).  In Fig.~\ref{fig:twocol1}(a) we only show objects
with $I_c>22.0$ mag and $V>22.0$ mag, because photometry of objects
brighter than these limits would be unreliable due to saturation or
detector non-linearity.
Out of 1237 objects with spectroscopic redshifts smaller than 4.5,
only two fall within our $z \sim 5$ colour selection criteria.  Both
of them have $z'$-band magnitude $\approx 22$ and their redshifts are
0.32 (ID=7066 by \citealt{tk04}) and 0.64 (ID=3249).  For the
J0053+1234 region, we show objects with published spectroscopic
redshift smaller than 4.5 with magenta squares in
Fig.~\ref{fig:twocol1}(b).  One sensible exception to this is CDFb-G5
at $z=4.486$, which satisfies our colour selection criteria
($V-I_c=2.70$, $I_c-z'=0.29$) and is shown with a filled green circle
as one of the spectroscopically confirmed LBGs at $z \sim 5$ (see
section 3.5).  In addition to this object, two objects at $z<4.5$ are
within the colour selection criteria for $z \sim 5$ LBGs.  One of them
is object id D0K 163 in \citet{coh99a}.  The redshift provided in
\citet{coh99a} is $z=0.580$ and is tagged with 'absorption only;
faint; id uncertain'.  It has a blue colour in $I_c-z'$ (0.07) in our
photometry.  Another object is CDFa-GD12 in \citet{steidel99}.  The
object is selected with their $z \sim 4$ LBG criteria (using $G$,
$\cal R$, $I$ filters) and the spectroscopic redshift is $z=3.469$.
Although the $I$-band magnitude listed in \citet{steidel99} is similar
to our $z'$-band magnitude ($z'=24.76$ mag), the distance between the
coordinate listed in \citet{steidel99} and ours are fairly large
(1.74\arcsec), unlike other objects (mostly $<1\arcsec$).

In \citet{i03} we defined the colour criteria for selection of
$z\sim5$ LBGs using the colours of galaxies with spectroscopic and
photometric redshifts in the HDF-N region, as well as model colour
tracks.  Since the publication of I03 the number of low-redshift
galaxies with spectroscopic identifications in the HDF-N region has
greatly increased, but, as described above, the cross-identification
of these spectroscopic samples with our LBG catalog shows that the
number of low-redshift galaxies that contaminate our colour selection
criteria is very small.

Since the magnitude range of our LBG candidates is fainter than most
of these spectroscopically identified low-redshift objects, the
objects identified as contaminants cannot be used to estimate the
fraction of contaminants in our sample.  Consequently, we need
resampling tests to estimate the contamination fraction as a function
of magnitude, as described in section 4.1.

Our HDF-N region contains the GOODS-N area, where deep HST/ACS and
Spitzer/IRAC and MIPS imaging have been made (\citealt{gia04a}). The
ACS images are in the $B_\mathrm{435}$, $V_\mathrm{606}$,
$I_\mathrm{775}$ and $z_\mathrm{850}$ filters and using
$V_\mathrm{606}$, $I_\mathrm{775}$ and $z_\mathrm{850}$, a $V$-dropout
LBG sample has been constructed \citep{rav06}.  However, it is not
straightforward to compare our sample with those based on the ACS
images. Since the filters used are different, especially
$V_\mathrm{606}$ which has a very broad bandpass (width is $\simeq
1500\mathrm{\AA}$), the colour tracks of the same galaxy SEDs with
GOODS/ACS filters in the two-colour diagram are quite different from
ours.

\begin{figure*}
\resizebox{!}{6cm}{
\includegraphics{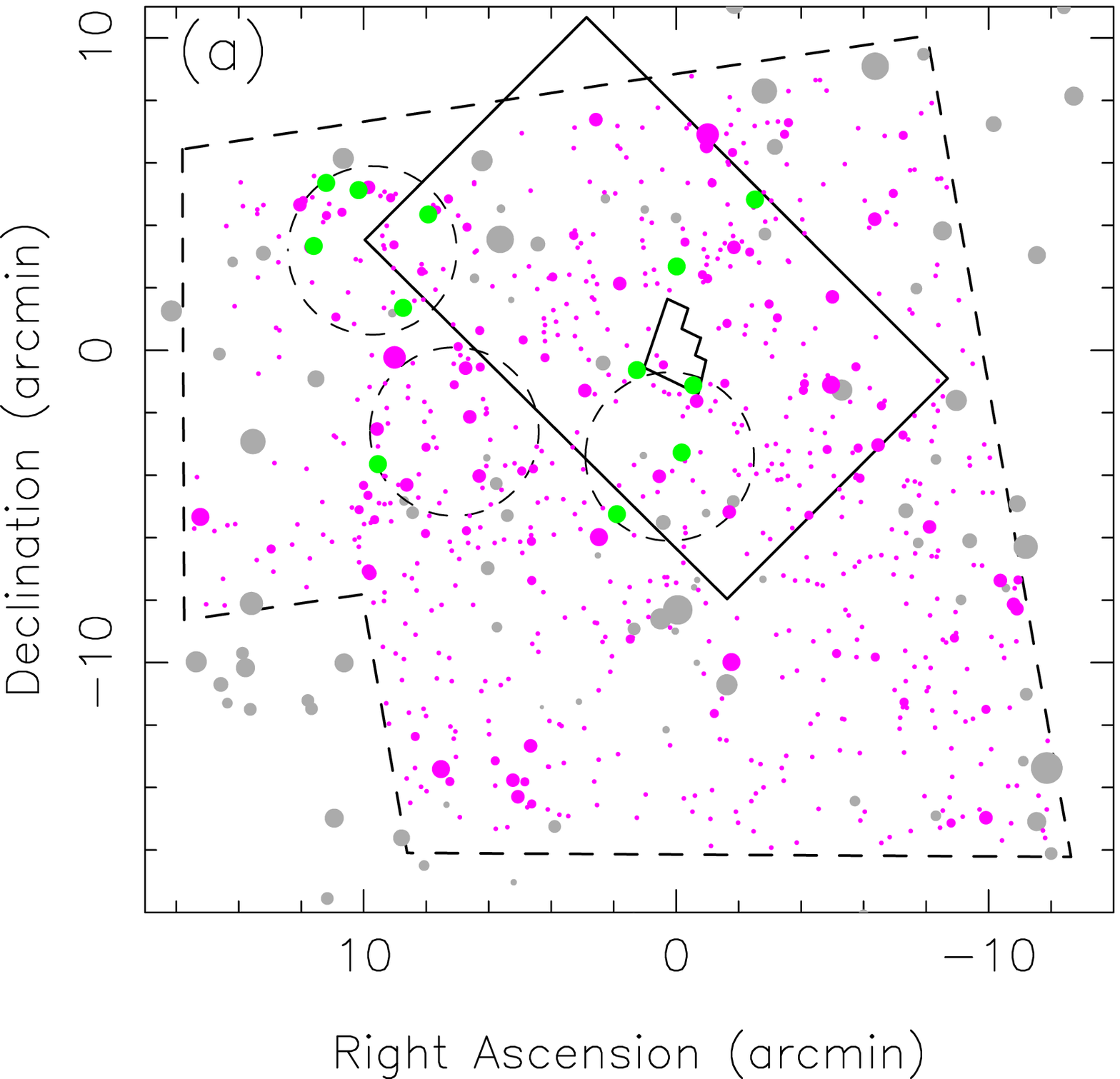}
\hspace{1cm}
\includegraphics{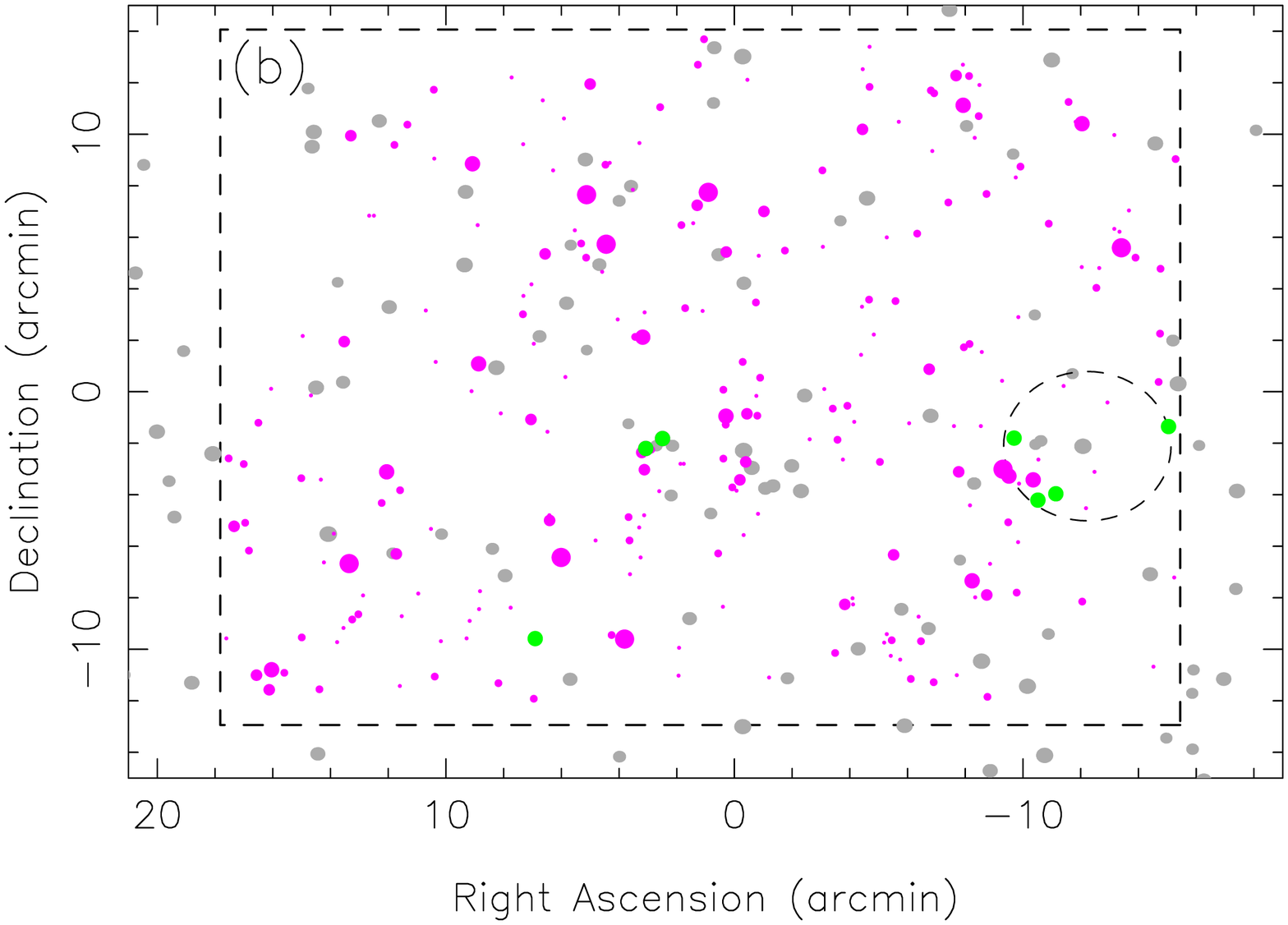}}
\caption{Sky distribution of LBG candidates 
in the HDF-N region (a) and in the J0053+1234 
region (b). Filled magenta circles  show the 
positions of LBG candidates. Symbol size indicates 
objects' brightness in a 0.5 mag step. 
Filled green circles are for objects which have been confirmed 
to be at $z \ga 4.5$ with optical spectroscopy. 
Large circles with dashed lines indicate the positions 
and areas of optical spectroscopy by us. 
Gray circles indicate the position of masks used 
to avoid the effects of bright stars. 
Rough boundaries of image areas used for selection of 
LBG candidates are shown with dashed lines. Note 
that not all of the area within these lines is used: 
due to the gaps between CCD chips, different position 
angles and dither scale lengths, some positions near 
boundaries have 
smaller effective exposure time, and such areas 
were not used for candidate selection.
Large and small areas drawn with solid lines in 
Fig.~\ref{fig:dist}(a) are 
the HST survey areas of the GOODS-N and the HDF-N, 
respectively.}
\label{fig:dist}
\end{figure*}

In Fig.~\ref{fig:dist} the sky distributions of LBG candidates in our
two fields are shown.  There is a clear fluctuation in surface density
of LBG candidates; some areas have many LBGs while others don't, and
these irregularity in the surface density is not caused by the masks
used to avoid effect of foreground bright objects such as galactic
stars and nearby galaxies. Fairly strong clustering signals have been
found with past wide-field LBG searches (e.g., \citealt{gia98};
\citealt{ade98}; \citealt{fou03}; \citealt{sdf6}; \citealt{ham06}). 
These figures also tell us that in small-area surveys such as the
HDF-N, the Hubble Ultra Deep Field and even GOODS, the effect of
field-to-field variance would be much more serious than for our data.
Detailed analyses of clustering properties of our LBG samples will be
presented in a forthcoming paper.

\subsection{Spectroscopic Identifications of Galaxies at $z\sim5$}

We carried out optical spectroscopic observations using a 
multi-object spectroscopy mode of the FOCAS \citep{kas02}
attached to the Subaru telescope. 
Until now, 8 objects in the HDF-N region and 4 objects in the 
J0053+1234 region 
have been identified to be in the redshift range between $z = 4.2$ and 5.2 
(\citealt{and04}; \citealt{and07}; including one AGN GOODS
J123647.96+620941.7).  No object with smaller redshift has been found
in our observations so far, although more than half of the observed
spectra have too low a signal-to-noise ratio to determine their
redshifts.  Four objects which have slightly redder $I_c-z'$ colours
than the colour selection criteria have been identified as Galactic
M-type stars \citep{and04}.

In addition, there are 8 objects in the HDF-N region and 2 objects in
J0053+1234 which have been spectroscopically identified to be at $4.5
< z < 5.5$ by other researchers.  Among them, 6 objects are selected
as $V$-dropout candidates in our images.  Tables
\ref{tbl:crossid_hdfn} and \ref{tbl:crossid_j0053} provide summaries
of the cross-identifications of these objects with previously
published spectroscopic redshifts.  In Fig.~\ref{fig:twocol1}, the
objects that have been selected as $V$-dropout candidates in our
images and that have spectroscopic confirmations are shown with filled
green circles (or upper arrows for objects not detected in $V$-band).
There are four objects in the HDF-N region which have been previously
reported to have spectroscopic redshifts larger than 4.5 and are
detected in our images but do not satisfy our colour selection
criteria (see table \ref{tbl:crossid_hdfn}).  Their positions in the
$V-I_c$ and $I_c-z'$ diagram are shown as open circles in
Fig.~\ref{fig:twocol1}.  The first of these, HDF-N 4-439.0 at
$z=4.54$, has $V-I_c=1.48$ and is slightly off from our colour
selection window.  Another object, F36219$-$1516, which has been
reported to be at $z=4.89$ by \citet{daw01}, has a very blue $V-I_c$
colour in our image ($V-I_c=0.67$).  Since the redshift determination
was made with solely a single emission line (identified as
Ly$\alpha$), we suggest that this may be a misidentification of a
lower redshift object.  F36376$-$1453 is another object with a single
emission line in the spectrum which was identified as Ly$\alpha$ at
$z=4.886$ by \citet{daw01}.  This object is very bright ($z'=21.97$)
and its colours in both $V-I_c$ and $I_c-z'$ are red ($V-I_c=3.25$,
$I_c-z'=0.78$ and this object is outside of plotted area of
Fig.~\ref{fig:twocol1}).  It has a round shape in our images in $I_c$
and $z'$-bands (FWHM$\sim 1.0''$).  The fourth object, GOODS
J123721.03+621502.1, was identified as an object at $z=4.76$ by
\citet{cow04} (no. 1173 in their table 1).  This object also has red
colours in $V-I_c$ and $I_c-z'$.  No information about the quality of
the spectrum is provided in \citet{cow04}.  On the other hand, one
object in the J0053+1234 region which is at $z=4.486$ (CDFb-G5 in
\citealt{steidel99}) and is within our selection window ($V-I_c=2.70$,
$I_c-z'=0.29$; see table \ref{tbl:crossid_j0053}), and it is shown as
an filled green circle in Fig.~\ref{fig:twocol1}(b).  In summary,
spectroscopic observations for our LBG candidates still cover only a
small fraction of the sample, but the results obtained so far are
encouraging.  Both the existence of objects at $z \sim 5$ within the
selection window and the small number of contaminants described in
section 3.4 strongly support the validity of our selection criteria.

\citet{cap04} analysed multicolour data including some of the same
Suprime-Cam data that we now re-processed and included in the present
work to construct our multi-band object catalog in the HDF-N region.
They also made a selection of $V$-dropout galaxies, derived their
number counts, and compared them with those in \citet{i03}. The number
counts in \citet{cap04} are just 13--21\% of those in this paper. They
claimed that our sample may be contaminated by $z \simeq 1$ galaxies
since our selection criteria are redder than theirs in $I_c-z'$
colour. However, their arguments are incorrect.  The colour criteria
used in \citet{cap04} are $V-I_c \ge 2.4$ and $V-I_c \ge 7(I_c-z') -
0.2$ and the latter criterion is slightly redder than ours
(in both \citet{i03} and the present paper we use $V-I_c > 7.0 (I_c -
z') + 0.15$ when on the AB normalization), contrary to their argument. 
Instead, the main cause of
the small number counts with their selection is that their first
criterion sets too strong a lower limit in $V-I_c$ colour and thus
misses many $z \simeq 4.5$ galaxies. As is shown in
Fig. \ref{fig:twocol1}, their criterion, $V-I_c
\ge 2.4$, misses many LBGs with high-$z$ spectroscopic identifications.

\subsection{UV colour distribution}

\begin{figure}
\resizebox{8cm}{!}{
\includegraphics{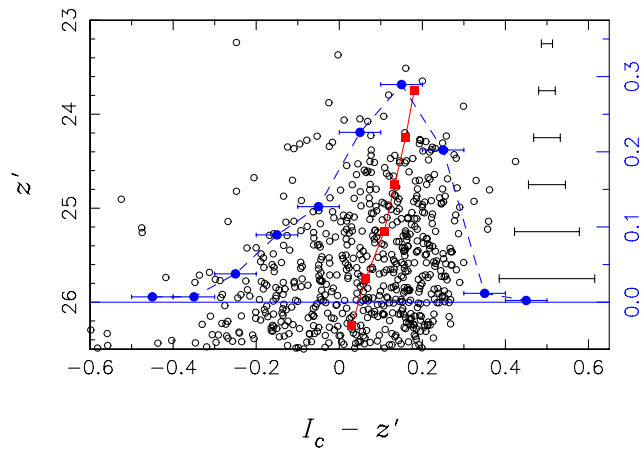}}
\caption{$z'$-band magnitude against $I_c-z'$ colour for 
LBG candidates at $z \sim 5$ in the HDF-N region. 
Open circles show magnitudes and colours of 
individual galaxies.
Typical errors in $I_c-z'$ colours 
are displayed in the right side of the panel. 
Solid squares show the median of $I_c-z'$ colours in 
a 0.5 mag step in $z'$-band magnitude. 
Filled circles indicate the distribution of fraction of 
$I_c-z'$ colours in an 0.1 dex step for objects 
with $23.0 < z' < 26.0$, and its scale is shown in 
the right side vertical axis.}
\label{fig:col_mag}
\end{figure}

If our LBG candidates are real objects at $z \sim 5$ then their
$I_c-z'$ colours represent rest-frame UV colours.  In
Fig.~\ref{fig:col_mag} we show the distribution of $I_c-z'$ colours
along apparent $z'$-band magnitude.  In this figure we also show the
distribution of $I_c-z'$ colours for objects with 
$23.0 < z' < 26.0$ as filled circles and the median colour values in 
different $z'$-band magnitudes (in a 0.5 mag step) as squares.  
Typical photometric errors in $I_c-z'$ colours are calculated from 
the simulations using artificial objects (described in section 4.1),
assuming $I_c-z'=0.07$ mag (from a model SED at $z=4.8$,
$E(B-V)=0.2$).  Because upper limits on $I_c-z'$ colour values are
imposed during the selection of LBGs, there are few objects with
$I_c-z' > 0.4$. This does not mean that all star-forming galaxies at
$z \sim 5$ have such blue UV colours. We do not know what fraction of
star-forming galaxies with red UV colours (probably due to heavy
attenuation by dust) at the target redshift are missed in our search;
at least in the redshift range $z = 2$--3 significant populations of
star-forming galaxies with redder rest-frame UV colour have been
detected (e.g, \citealt{fs04}; \citealt{dad04}).  Application of
similar methods at higher redshifts will be an important subject of
future space-based infrared missions, but for now we must press on
with UV-selected samples.

In the median values of $I_c-z'$ colours there is a weak trend in that
objects bright in $z'$-band have redder $I_c-z'$ colours.  In other
words, the ratio of the number of blue (e.g., $I_c-z' < 0$) objects to
the number of red objects is smaller for bright objects than for
fainter ones.  Since the upper limit on $I_c-z'$ colour in our
selection criteria depends on $V-I_c$ colour, the selection effect is
not straightforward in this figure.  However, since there is no
restriction for the selection of objects with blue $I_c-z'$ colours,
the absence of bright objects with blue rest-frame UV colour must be
real.  There is no selection bias due to the limiting magnitude in
$I_c$-band, since this is $I_c=26.86$ mag in the HDF-N region.  The
red colour of UV continua can be attributed to various origins,
including heavier dust attenuation, older stellar populations, and
weaker Lyman $\alpha$ emission (for objects at $z>4.8$ the $I_c$
filter covers the rest-frame Lyman $\alpha$ wavelength).  We will
briefly discuss this dependence of the UV colour distribution on UV
luminosity in section 5.2.2.

\section{UV Luminosity Function at $z \sim 5$}

\subsection{Correction for Incompleteness and Contamination}

The procedure that we use to derive the UV luminosity function follows
that described in our previous paper \citep{i03} which was based on
shallower data for the HDF-N region only.  To estimate the volumes
sampled with our adopted colour criteria, we execute a large set of
tests in which we use artificial galaxies which mimic the size
distribution of real galaxies in our sample.  We are able to derive
the completeness of our survey for objects with specific magnitudes
and colours by counting the number of artificial objects which have
been recovered by the object detection and colour selection
procedures.  In a single test run we generate 25,000 artificial
objects for each bin of 0.5 in magnitude, 0.1 in redshift, and
reddening by dust (five different degree of reddening from
$E(B-V)=0.0$ to 0.4 with steps of 0.1). Colours are assigned based on
the model spectra of star-forming galaxies generated with PEGASE.2
(see Fig.~\ref{fig:twocol0}), \citet{cal00} extinction law, and a
prescription for IGM attenuation by \citet{ino05}.  These objects are
then put into random positions in the $V$, $I_c$ and $z'$ images. The
methods of object detection and sample selection according to the
colour criteria are the same as those for real objects, described in
sections 3.3 and 3.4.

\begin{figure}
\resizebox{8cm}{!}{
\includegraphics{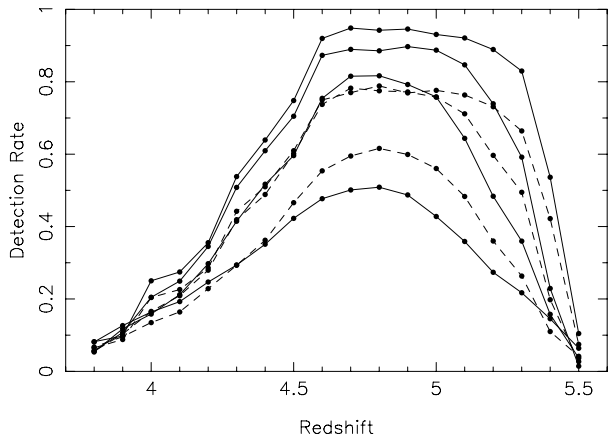}}
\caption{
Detection rates of the galaxies for the HDF-N region (solid lines) and
the J0053+1234 region (dashed lines), as a function of $z'$-band
magnitude and redshift. The lines represent $z'$-band magnitude of
23.0--23.5, 24.0--24.5, 25.0--25.5 and 26.0--26.5 (the last bin for
the HDF-N region only).  }
\label{fig:selfunc}
\end{figure}

In Fig.~\ref{fig:selfunc} we show the completeness of our survey as a
function of redshift and $z'$-band magnitude. Results for five
different amounts of reddening were averaged with weights estimated
from the UV colour distribution of real galaxies (section 3.6).  The
maximum completeness is achieved for brightest (23.0--23.5 mag)
objects at redshift 4.5--5.0 in the HDF-N region.  We note that there
is a significant ($\ga 10$\%) detection rate for objects at $z \sim
4$.  It suggests that in our sample of $z\sim 5$ LBG candidates there
should be a small number of objects at such lower redshift ranges.  So
far we have detected one object at $z=4.26$ in the optical
spectroscopy of the J0053+1234 region (table \ref{tbl:crossid_j0053};
\citealt{and07}).  The existence of such relatively low redshift
objects does not have a serious effect in deriving the UVLF at $z \sim
5$, since we properly treat them by calculating the survey volume
using these selection functions.
The detection rates in the J0053+1234 region are lower than those in
the HDF-N region at the same magnitude ranges, and they are similar to
those for the data used in \cite{i03}, which were $\sim 0.8$ mag
shallower in $z'$-band than the present HDF-N data.

A weighted average value of redshift derived from this completeness
distribution is 4.8 for both the HDF-N and the J0053+1234 regions.
Hereafter we use this value as a typical redshift in the presentation
of the UV luminosity function.  From these completeness estimates for
different magnitudes, redshifts and amount of dust reddening, we
calculated the {\it effective volumes} \citep{steidel99} of our
surveys, which are listed in Tables~\ref{tbl:hdfuvlf} and
\ref{tbl:j0053uvlf}.

As we did in \citealt{i03}, we use a resampling method to estimate the
rate of contamination by galactic stars and galaxies at intermediate
redshifts.  We perturbed the colours of detected objects which lie
outside of the colour selection criteria by adding random errors based
on the error measurements in photometry. Then we counted the number of
objects which fell into the colour selection criteria after the
addition of these random errors.  For each bin of 0.5 magnitude these
resampling tests were executed 1,000 times, and average numbers of
interlopers were derived. These numbers are listed in
Table~\ref{tbl:hdfuvlf} and \ref{tbl:j0053uvlf} as
$N_\mathrm{int}$. 
\footnote{We also tested a case in which we perturbed 
the colours of objects within the colour selection area and 
count the number of objects which get out from the area, and 
confirmed that the changes in the UVLF are smaller than or 
comparable to the size of overall errors. See section 4.1 
of \citet{i03} for more deitals.}

\begin{figure*}
\resizebox{12cm}{!}{
\includegraphics{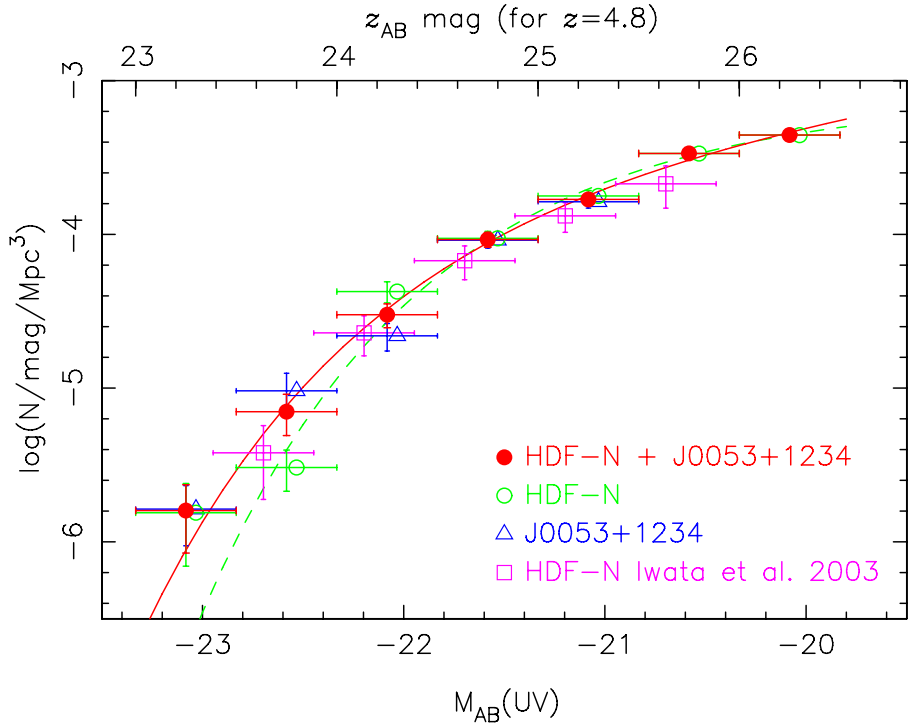}}
\caption{
UV luminosity functions (UVLFs) at $z \sim 5$.  Open circles and open
triangles represent UVLFs of LBGs at $z \sim 5$ in the HDF-N region
and the J0053+1234 region, respectively.  Filled circles show the UVLF
derived by combining results of the two fields.  The data points of
UVLFs based on single fields are shifted +0.05 mag for clarity.  The
solid line represents the Schechter function fitted to the combined
UVLF, and the dashed line is a fitting for the UVLF in the HDF-N.
Open squares are the UVLF in the HDF-N region determined with previous
shallower data (I03).  }
\label{fig:lf1a}
\end{figure*}

We compute the UVLF as 
\[
 \Phi(m) = 2 \times (N(m) - N_\mathrm{int}(m)) / V_{\mathrm{eff}}(m),
\]
where $N(m)$ represents the number of LBG candidates detected within
the magnitude bin of $m \pm 0.25$ mag, and $V_{\mathrm{eff}}(m)$ 
is the effective volume.  The multiplication by 2 is
required to obtain number densities per magnitude, since we take 0.5
mag bins in our sample.  The UVLFs derived for the HDF-N region and
the J0053+1234 region are listed in Tables~\ref{tbl:hdfuvlf} and
\ref{tbl:j0053uvlf}, respectively, and they are shown in
Fig.~\ref{fig:lf1a} with open circles (HDF-N) and open triangles
(J0053+1234).  The error bars indicate 1$\sigma$ error estimates
including statistical (Poisson) noise, photometric errors and
uncertainties in contaminations and completeness.  In converting from
apparent $z'$-band magnitude to absolute magnitude we assume a
luminosity distance of an object at $z=4.8$ under the adopted
cosmology ($M_{\mathrm UV}=m_{z'}-46.33$, $\Omega_M=0.3$,
$\Omega_\Lambda=0.7$, $H_0=70$ km/s/Mpc).  In \citet{i03} we made a
correction for luminosity in the conversion from $I_c$-band apparent
magnitude to rest-frame luminosity at 1700$\mathrm \AA$, assuming the
SED of a model star-forming galaxy.  In the present analysis we do not
make such a correction in the conversion from magnitudes in $z'$-band
(the effective wavelength is $9086 \mathrm{\AA}$), and thus the LFs we
present are at $\approx 1570 \mathrm{\AA}$.  

\begin{figure*}
\resizebox{12cm}{!}{
\includegraphics{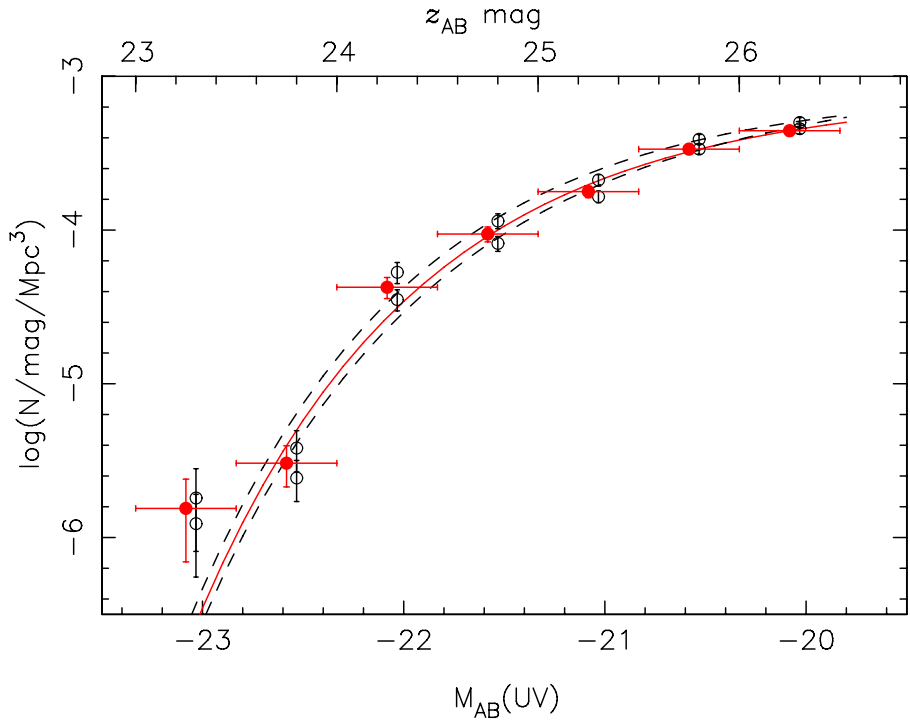}}
\caption{The UVLF at $z \sim 5$ for the HDF-N region 
(same as the UVLF with open circles in Fig.~\ref{fig:lf1a}) calculated
using the distribution of dust attenuation estimated from $I_c-z'$
colours (Fig.~\ref{fig:col_mag}) is compared with UVLFs derived based
on two extreme assumptions about the dust attenuation of our sample
LBGs (no attenuation and $E(B-V)=0.4$ for all objects), shown with
open circles. LFs with open circles are shifted +0.05 mag for
clarity. See section 4.1 for details.  }
\label{fig:lf1.5}
\end{figure*}

In the calculation of selection functions and effective volumes we
assume a dust attenuation distribution from $E(B-V)=0.0$ to 0.4 based
on the $I_c-z'$ colour distribution. In order to examine the
robustness of the LF against this assumption about dust attenuation,
we also calculated effective volumes with two extreme cases in dust
attenuation: one case is that using just a model without dust
attenuation ($E(B-V)=0.0$) and the other is only using a model with
relatively large dust attenuation ($E(B-V)=0.4$). In these extreme
cases the shape of selection function along redshift changes from that
shown in Fig.~\ref{fig:selfunc}, because the redshift range covered
with our selection criteria changes when the assumed model is
different. However, we find that the differences in the resulting LF
--- even with these extreme cases --- are small, as shown in
Fig.~\ref{fig:lf1.5}.  In this figure the UVLFs in the HDF-N region
using effective volumes calculated based on these extreme assumptions
are compared with the UVLF computed using the fiducial dust
attenuation distribution.  Since the extreme assumptions in these
tests (all objects are free from dust or all objects have fairly large
dust attenuation of $E(B-V)=0.4$) are unlikely, we can say that our
UVLF is robust against the assumption of model colours and
distribution of amount of dust attenuation.

The modification of the selection criteria should also alter the
effective volume.  We examined the change of effective volumes and
luminosity function by means of a test in which the colour selection
criteria were shifted in both $V-I_c$ and $I_c-z'$ with 0.05--0.15
magnitude, and derived luminosity functions in the same way as we did
for the original selection criteria.  The UVLFs obtained by these
tests show differences smaller than the error bars of our LF shown in
Fig.~\ref{fig:lf1a}.  With these tests we verified that our UVLF
results are robust against slight changes to the colour selection
criteria.

\subsection{Combination of the UVLFs in the two survey fields and Schechter
  function fitting}

Since the J0053+1234 sample is shallower than the HDF-N sample, the
UVLF of the J0053+1234 region is limited to $z' < 25.5$ mag.  As shown
in Fig.~\ref{fig:lf1a}, the number densities of these two fields agree
fairly well, although in two bright magnitude bins ($23.5 < z' < 24.0$
and $24.0 < z' < 24.5$) the discrepancy between results from the two
fields is large enough that it exceeds the size of the 1-$\sigma$
error bars.  This is probably due to the small number of objects
(5--35) in these magnitude bins, and it might also indicate the
existence of field-to-field variations, the effect of which is not
included in the error estimates.

We averaged the number densities of the two fields by applying weights
based on their survey areas (for magnitude range brighter than $z' =
25.5$), and derived the final UVLF with the present data set. This
UVLF is listed in Table~\ref{tbl:uvlfmean} and is shown as filled
circles in Fig.~\ref{fig:lf1a}. Hereafter we call this the final UVLF.

\begin{figure*}
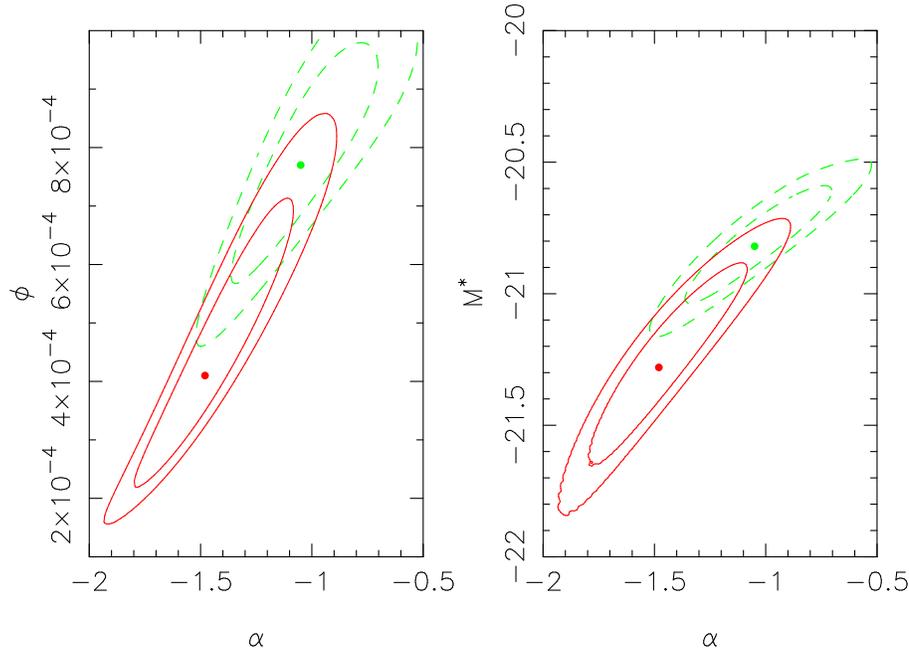

\resizebox{12cm}{!}{
\includegraphics{fig09a.eps}
\includegraphics{fig09b.eps}}
\caption{
Confidence contours of Schechter function parameters.  Solid lines
show areas show the 68\% (inner) and 95\% (outer) confidence levels
for the 'final' UVLF which combines the results of the two fields.
Dashed lines are the same as the solid lines, but for the (deeper)
HDF-N data only.  Small points show the values of the best-fitted
parameters.  }
\label{fig:lfconf1}
\end{figure*}

The results of fitting the UVLFs with the Schechter function
\citep{sch76} are also shown in Fig.~\ref{fig:lf1a}. The solid line is
a function fitted to the final UVLF, and the dashed line is for the
UVLF based solely on the deeper HDF-N region data.  The best-fitted
parameter values for the final UVLF and for the LF from the HDF-N
region are summarized in Table~\ref{tbl:uvlfpar}.  We note that the $z
\sim 5$ Schechter parameter fit given in \citet{kdf3} is based on a 
(slightly earlier) version of the HDF-N region UVLF presented here.
Fig.~\ref{fig:lfconf1} also shows the 68\% and 95\% confidence level
error contours calculated via Monte-Carlo resampling of the UVLFs. For
each point of the observed UVLFs we add a random error according to
its uncertainty (assuming Gaussian distribution), and execute the
Schechter function fitting to this perturbed UVLF and obtain a
$\chi^2$ value. We repeat this test 1,000 times and derive the region
in parameter space that contains 68\% or 95\% of these resampling
results.

Although the number densities in the combined data and those in the
HDF-N region are fairly close in the fainter part of the LF ($24.0 <
z' < 25.5$), the best-fitted Schechter function parameters are
significantly different. This difference is caused by relatively large
difference in the number density in the bright $23.5 < z' < 24.0$
magnitude range.  Even the faint-end slope $\alpha$ is affected by the
change in the LF's bright part, due to changes in $\phi^\ast$ and
$M^\ast$.  The large field-to-field variation in the bright end of the
UVLF between our two survey areas (both fields have areas larger than
500 arcmin$^2$) clearly indicates the necessity of a large survey area
and multiple survey fields. A study based on a single field
observation may be affected by large-scale structure in the galaxy
distribution. Moreover, if the survey area is small, it is very likely
that the survey fails to detect bright and rare objects and so
underestimates their number density.  We also note that the Schechter
function parameters are not a good way to represent the properties of
observed LFs unless an observation reaches far deeper than $M^\ast$.
A comparison of LFs at different epochs or in different environments
should be cautious of uncertainties and degeneracy between parameters
(as illustrated in Fig.~\ref{fig:lfconf1}).  The direct comparison of
number densities at in a fixed luminosity range is a better way to
compare LFs.

\subsection{Comparisons with UVLFs in previous studies of galaxies at $z \sim 5$}

\subsubsection{Comparison with the UVLF in Iwata et al. (2003)}

The new UVLFs at $z \sim 5$ agree well with our previous result
\citep{i03} based on shallower data in the HDF-N region, which is
indicated as open squares in Fig.~\ref{fig:lf1a}.  Note that because
in \citet{i03} we used the Vega-based magnitude system and the
fiducial redshift was $z=5.0$, the magnitude bins are slightly
different from our present results.

\subsubsection{Comparison with the UVLF of FORS Deep Field}

\begin{figure*}
\resizebox{12cm}{!}{
\includegraphics{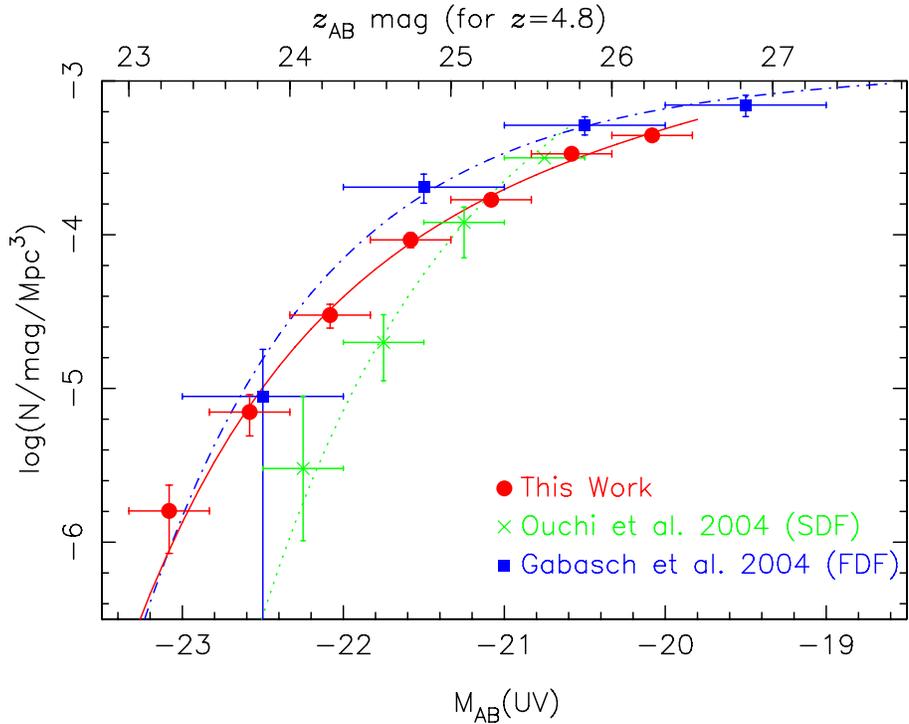}}
\caption{
UV luminosity functions at $z \sim 5$. 
Filled circles show our UVLF derived by combining 
results of the two fields. This is the same UVLF as in Fig.~\ref{fig:lf1a}. 
The solid line represents the Schechter 
function fitted to this combined UVLF. 
Solid squares represent the rest-frame 1500\AA\ luminosity function 
derived by \citet{gab04a} 
from their $I$-band selected
photometric redshift sample with estimated redshifts 
at $4.01 < z < 5.01$,
and the dot-dashed line is a Schechter function fitted 
to it.
The crosses are the UVLF for LBGs at $z\sim5$ ($V$-band 
drop LBGs) by \citet{sdf5} 
and the dotted 
line represents the Schechter function fitted to 
their data with the faint-end slope fixed to $\alpha = -1.6$.
}
\label{fig:lf1b}
\end{figure*}

In Fig.~\ref{fig:lf1b} we compare our UVLF with the one at 1500 \AA\
which is based on the photometric redshift sample selected in $I$-band
by \citet{gab04a}. These authors used 150 galaxies in the FORS Deep 
Field (FDF) with $I$-band 
magnitudes brighter than 26.8 mag which have estimated redshifts
between 4.01 and 5.01. While the depth of their sample is comparable
to ours\footnote{We should note that the $g$- and $R$- band images
used in \citet{gab04a} have depths comparable to that in $I$-band and
may not be deep enough to detect the spectral break reliably, while
their $B$-band image is $\sim0.8$ mag deeper than $I$-band.}, the
number of objects is small, mainly due to their small survey field
area ($7' \times 7'$).  The overall shape of the UVLF derived by
\citet{gab04a} is similar to ours. The number density of faint
galaxies in our UVLF at $z \sim 5$ is slightly smaller, as has been
pointed out by \citet{gab04a} who used our previous UVLF for the
comparison.  Since the redshift range of galaxies in \citet{gab04a} is
smaller than ours, this trend might indicate a real number density
evolution of UV-selected star-forming galaxies (see section 5.1 for
discussion of the differential UVLF evolution), 
although the difference in the methods of sample selection and possible 
field-to-field variation prevent us to make further discussions.

\subsubsection{Comparison with the result of Subaru Deep Survey}

In Fig.~\ref{fig:lf1b} we show with crosses the UVLF of LBGs at $z
\sim 4.7$ derived by \citet{sdf5} in the Subaru Deep Field and show 
their fitted Schechter function using the dotted line.  Note that
their faint-end slope $\alpha$ was fixed to $-1.6$.  As described in
\citet{sdf5}, there is a discrepancy in the bright part
($M_\mathrm{UV} < -21.5$ mag) between our previous UVLF \citep{i03}
and their result.  The discrepancy still remains when our updated data
is compared with the \citet{sdf5} UVLF, and this seems somewhat
unnatural, since the field-to-field variance between our two survey
fields (at $M_\mathrm{UV}<-22$) is more than three times smaller than
the difference between our UVLFs and the one by \citet{sdf5}.
Recently, \citet{yoshida06} derived an updated UVLF of $z \sim 5$ LBGs
in the same Subaru Deep Field using a deeper and $\approx$160
arcmin$^2$ wider image data set.  Their filter set, procedures for
reducing data, selecting LBG candidates, and calculating the UVLF are
basically the same as those in \citet{sdf5}. Consequently, their
results agree with those by \citet{sdf5} and thus the number densities
in their LF's bright part remains significantly smaller than ours.

\begin{figure*}
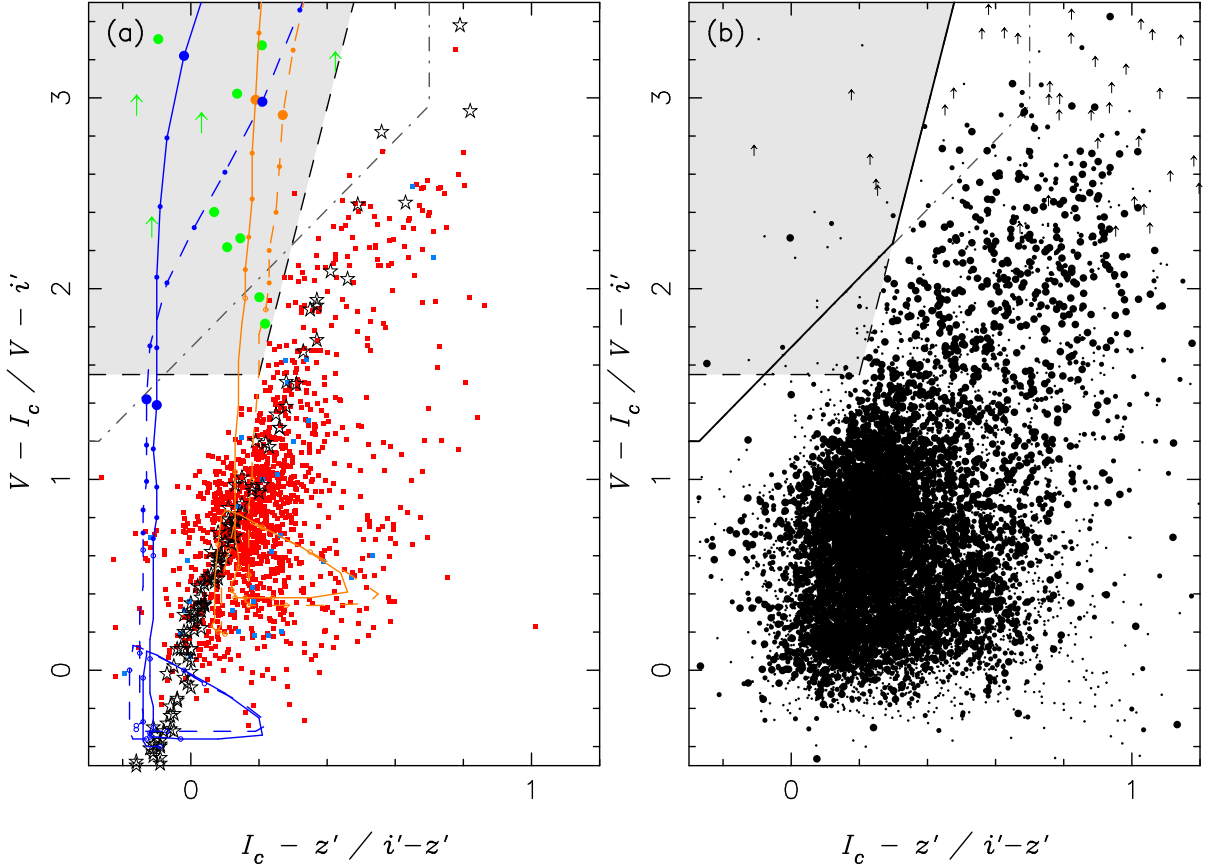

\resizebox{16cm}{!}{
\includegraphics{fig11a.eps}
\includegraphics{fig11b.eps}}
\caption{
$V-I_c$ and $I_c-z'$ two colour diagrams. 
(a): 
Red and blue squares are galaxies at $z < 4.0$ in the 
HDF-N region, same as in Fig.~\ref{fig:twocol1}a.
Green circles are spectroscopically confirmed objects at 
$z > 4.5$ in the HDF-N.
For objects not detected in $V$-band, arrows are used to 
show the 3-$\sigma$ lower limits of $V-I_c$ colours. 
The dashed line and shaded area stands for colour selection 
criteria adopted in this study for the selection of Lyman 
break galaxies at $z\sim 5$. 
The colour selection criteria of Ouchi et al. (2004a) (used with 
$V-i'$ and $i'-z'$ colours) are displayed with a dot-dashed line.
Blue lines are colour tracks of a star-forming galaxy 
without dust attenuation, and lines in orange 
are for dust attenuation of $E(B-V)=0.4$. 
Solid lines show colours in $V-I_c$ and $I_c-z'$, and 
dashed lines are in $V-i'$ and $i'-z'$.
Star symbols indicate the $V-I_c$ and $I_c-z'$ colours of 
A0--M9 stars calculated based on
the library by Pickles (1998). 
(b): same as (a), but shows objects with 
$23.0 < z' < 24.5$  mag 
detected in our $V$, $I_c$ and $z'$-band images 
for the HDF-N region. 
Sizes of symbols indicate object brightness in 0.5 mag 
bins, with the largest symbols representing objects with 
$23.0 < z' < 23.5$ mag. 
The thick solid line represents the modified colour selection 
criteria used for the test described in section 4.3.3., 
which consist of the colour selection 
criteria of Ouchi et al. (2004a) for $i'-z' < 0.3$, 
and for $i'-z'>0.3$ follows our colour selection criterion. 
}
\label{fig:twocol2}
\end{figure*}

We further investigated the origin of the difference between our
results and those by \citet{sdf5}.  One possible cause of this
discrepancy is the difference in colour selection criteria.  However,
we must take care here because the filters used by us and those used
by \citet{sdf5} are different: \citet{sdf5} used $i'$ while we used
$I_c$, and since the $i'$-band filter has transmission that is shifted
slightly to a shorter wavelength with respect to the $I_c$-band (see
Fig.~\ref{fig:filters}), the comparison of sample selection criteria
is not straightforward.  We illustrate this problem as follows.  In
Fig.~\ref{fig:twocol2} we use a dot-dashed line to show the $V-i'$ and
$i'-z'$ colour criteria used in \citet{sdf5}.  Fig.~\ref{fig:twocol2}
also shows the redshift tracks of star-forming galaxies in the colour
pair $V-I_c$ and $I_c-z'$ (solid lines) and in the colour pair $V-i'$
and $i'-z'$ (dashed lines). Since the effective wavelength of
$i'$-band is slightly shorter than that of $I_c$-band, the $i'-z'$
colour becomes red at relatively low redshift (at $z \sim 4.7$).
If we were to apply the \citet{sdf5} criteria directly to our $V-I_c$
and $I_c-z'$ colours, then the sample would be contaminated by many
low-$z$ galaxies and galactic stars with $V-I_c \ga 2.4$ and $I_c-z'
\ga 0.4$, as shown with red squares and star symbols in
Fig.~\ref{fig:twocol2}(a). However, the selection criteria adopted by
\citet{sdf5} avoid including these objects because the $i'-z'$ colours
of these objects are $\sim0.2$ mag redder than their $I_c-z'$ colours.
This illustrates the difficulty in making a direct comparison of the
colour selection criteria of the two surveys.

\begin{figure*}
\resizebox{12cm}{!}{
\includegraphics{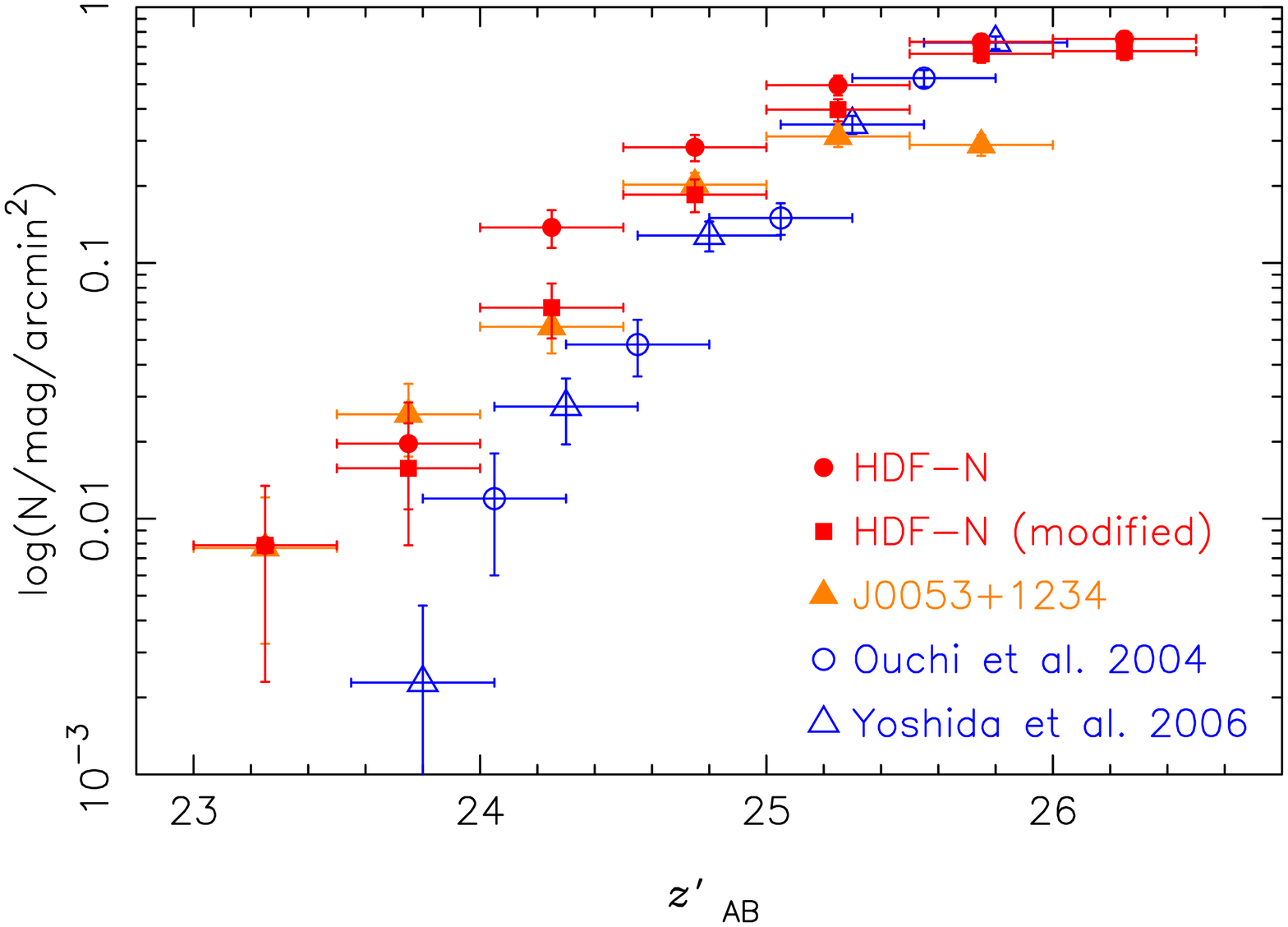}}
\caption{The $z'$-band number count of LBGs at $z \sim 5$. 
Filled circles and triangles are those for the HDF-N and 
J0053+1234 regions, respectively, based on the sample 
selected using our colour criteria. 
Number counts of $V$-drop LBGs in SDF are shown 
with open circles \citep{sdf5} and open triangles 
\citep{yoshida06}.
Filled squares show the number counts of LBGs in the 
HDF-N selected using the modified colour criteria 
based on those used in Ouchi et al.(2004). 
See text for details. 
No correction for incompleteness has been made. 
Vertical error bars show simple Poisson errors.
}
\label{fig:numcount}
\end{figure*}

Thus, in order to roughly estimate the effect of the difference
between colour selection criteria, we adopt modified criteria which
consist of those used by \citet{sdf5} at $I_c-z' < 0.3$ and those used
by us at $I_c-z' \geq 0.3$.  These modified criteria are shown with a
thick solid line in Fig.~\ref{fig:twocol2}(b); they approximately
reproduce the \citet{sdf5} criteria as applied to our $V-I_c$,
$I_c-z'$ data.  In Fig.~\ref{fig:numcount} we show $z'$-band number
counts of LBGs in our survey fields, using both our criteria and the
modified criteria of \citet{sdf5} and compare them with the number
counts given by \citet{sdf5} and \citet{yoshida06}.  
The change of colour selection criteria
reduces the number of LBGs in the HDF-N region, but there remains a
clear discrepancy between our number counts and those in the Subaru
Deep Field (\citealt{sdf5} and \citealt{yoshida06}).  In the 875
arcmin$^2$ of \citet{yoshida06} there is only one object with $z' <
24.05$ mag and only 12 objects with $z'$-band magnitude between 24.05
and 24.55,
while there are 7 objects with $z' < 24.0$ mag and 35 objects with
$24.0 < z' < 24.5$ mag in the HDF-N region when we use our own
selection criteria.  As shown in Fig.~\ref{fig:twocol2}(b), when we
adopt the modified colour criteria, six out of 7 LBG candidates with
$23.0 < z' < 24.0$ remain within the selection area.  In $24.0 < z' <
24.5$ the number of objects becomes 17, about half of our original
sample, but this corresponds to $\approx29$ for the survey area of
\citet{yoshida06} and so represents a surface number density more than 
two times larger than theirs.  Thus, the modification in the colour
selection criteria is unable to explain the differences between the
number densities in our final UVLF and those in the Subaru Deep Field,
which are 2--5 times larger than the 1$\sigma$ errors estimated for
our UVLF.

In addition to the change in the number density of objects, the change
of colour selection criteria reduces the effective volume. These two
effects counteract each other and, as a result, the luminosity
function would remain relatively unchanged.  Because of the difference
in the filter sets, we cannot calculate the effective volume in the
case of applying the colour selection criteria of \citet{sdf5} to our
sample. We are thus unable to calculate the UVLF with such criteria
and so must remain focused on comparing raw number counts.  However,
as we have mentioned in section 4.1, slight changes to the colour
selection criteria do not change the shape of the UVLF.

Although the number of objects in our sample which were observed with
optical spectroscopy is still limited, 6 out of 8 objects 
with $z'$-band magnitude between 24.0 and 24.5 mag in the HDF-N
region for which we obtained spectroscopy are confirmed to be at 
$z > 4.5$ 
(including one AGN; \citealt{and04}).  The redshifts of the remaining
two objects are unidentified due to the absence of clear spectral
features.  Thus, we believe that most of bright objects in our sample
are genuine LBGs at $z \sim 5$.  Indeed, if the success rate of
spectroscopic confirmation (6/8) for bright LBGs is representative of
that for the whole of our sample, it implies that at least 26 among
the 35 HDF-N-region galaxies with $z' = 24.0$--24.5 are at $z > 4.5$.
This number density of 26/505.85=0.051 arcmin$^{-2}$ well exceeds the
number density of candidates in the same magnitude range (12/875=0.013
N/arcmin$^{-2}$) in \citet{yoshida06}. Consequently, it is possible
that \citet{sdf5} and \citet{yoshida06} may miss significant numbers
of luminous high-$z$ objects unless the field-to-field variance
between the HDF-N region and SDF is far larger than expected.
However, such large field-to-field variance seems unrealistic
considering the spatial distribution of LBG candidates in our survey
fields and the similarity of the UVLF results in our two fields.

Thus, although the cause of this discrepancy remains an open question,
because of the comparisons and considerations described above, we
think our UVLF is robust.  Hereafter we use our final LF result as a
representative $z \sim 5$ UVLF to compare with UVLFs at lower redshift
ranges.

\subsection{Comparisons with results of numerical simulations}

Recent development of computer resources and numerical techniques
makes comparisons of observational results with results from numerical
simulations interesting since such comparisons make it possible to
investigate the physics behind the observed properties of galaxies.
\citet{ngt06} executed cosmological SPH simulations, 
calculated SEDs for galaxies in their simulations using the population
synthesis code of \citet{bc03}, and then applied LBG colour selection
criteria to create a simulated sample of LBGs at $z=$4--6.  They use
this sample to compare UVLFs in their numerical simulations with ours
(UVLF for the HDF-N region) and with \citet{sdf5}. Since the
prescriptions of \citet{ngt06} do not include attenuation by dust,
their UVLFs can be shifted along the luminosity axis by varying the
assumed amount of uniform dust attenuation.  The UVLF of $z=5$ LBGs in
their simulations can be made to agree with observed UVLFs (both ours
and that by \citealt{sdf5}) by varying dust attenuation between
$E(B-V)=0.15$ and $0.3$ (using the attenuation law by
\citealt{cal00}). \citet{fin06} applied dust attenuation based on a
correlation between the metallicity and dust attenuation inferred from
the large sample of nearby galaxies taken from the Sloan Digital Sky
Survey. They only show galaxies at $z=4$ in their numerical
simulations, and their UVLF is broadly consistent with observational
results at $z \sim 4$ such as \citet{kdf2}, \citet{sdf5} and
\citet{gab04a}.  Although the UVLFs found in numerical simulations by
\citet{ngt06} and \citet{fin06} show fairly steep slopes, 
$\alpha \ga -1.7$, such slopes are estimated from magnitude ranges
fainter than the limiting magnitudes of current deep observations ($m
\ga 27$).  Thus, in summary, no serious conflict with observed results
has been found in these numerical simulations. However, we should note
that there are free parameters in models used in these simulations
which are not strongly constrained by physics in a self-consistent way
(e.g., amount of dust attenuation, effects of feedback from star
formation and AGN), and it is still not possible to discriminate the
``best'' observed UVLF through the comparisons with results of such
numerical simulations.  One interesting prediction common in these two
simulations is that UV-luminous objects should have higher stellar
mass than low-luminosity ones.  This trend may be related to the
differential evolution of galaxies we describe in section 5, and this
prediction is testable with deep infrared observations of LBGs at $z\ga 4$.

\section{Differential Evolution of UV Luminosity Function from $z \sim 5$ to 3}

\subsection{Comparison with the $z \sim 4$ and 3 UVLFs by 
Sawicki and Thompson (2006)}

Sawicki and Thompson \citeyearpar{kdf1, kdf2} constructed
ultra-deep samples of UV-selected star-forming galaxies at $1.7 <z< 4
$ using the very same $U_n G {\cal R} I$ filters and colour selection
criteria as those used at brighter magnitudes by Steidel and his
collaborators.  Using these very deep samples, \citet{kdf2} derived
the UVLFs reaching far below $L^*_{z=3}$.  Specifically, at $z = 3$
their limiting magnitude is $M_{lim} \sim -18.5$ mag and $M_{lim}$ is
about $-19.0$ mag at $z \sim 4$.  These depths are about 1.5
magnitudes fainter than previous large-area studies (e.g.,
\citealt{steidel99}), but the results are highly reliable because of
the use of the selection technique {\it identical} (down to the same
filter set and color-color selection criteria) to that in the Steidel
et al.\ work.  Comparisons of our new $z \sim 5$ UVLFs with these deep
UVLFs at lower redshifts can lead to crucial insights regarding the
evolution of UV-selected star-forming galaxies during early epochs in
the history of galaxy formation.

\begin{figure*}
\resizebox{12cm}{!}{
\includegraphics{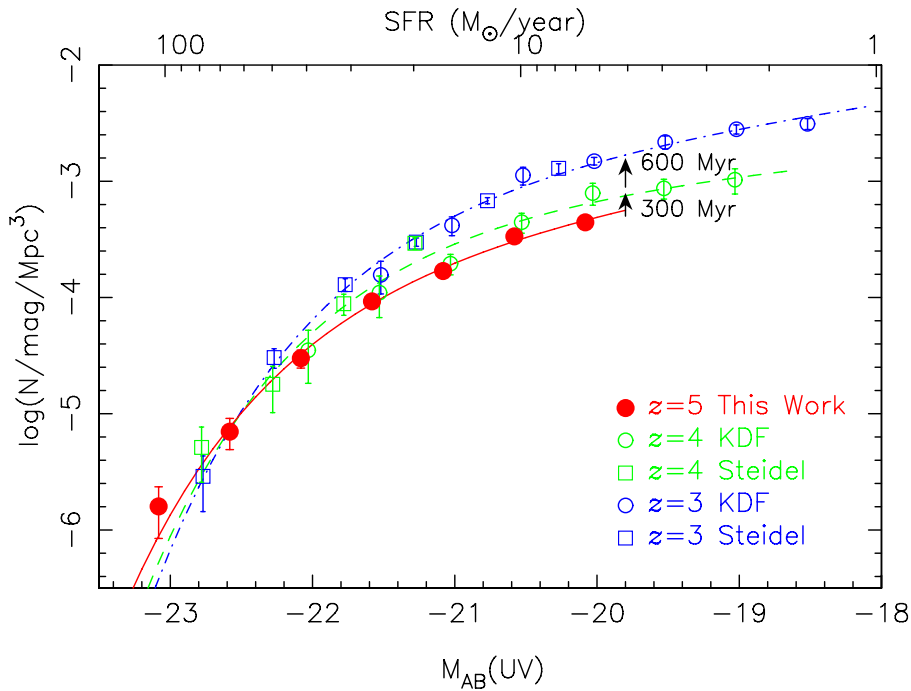}}
\caption{
UV luminosity function (UVLF) at $z \sim 5$ 
(solid circles and solid line) in this study 
and those at $z \sim 4$ 
(dashed line) and at $z \sim 3$ (dot-dashed 
line) by Sawicki and Thompson (2006a). 
}
\label{fig:lf2}
\end{figure*}

In Fig.~\ref{fig:lf2} we show our new $z \sim 5$ UVLF (the final
weight-averaged LF, as described in sec. 4.2) along with the $z\sim 4$
and $z \sim 3$ UVLFs from \citet{kdf2}.  In \citealt{i03} we compared
the previous, shallower version of our $z \sim 5$ UVLFs with $z=3$--4
UVLFs by \citet{steidel99} and argued that there is no significant
difference between UVLFs from $z=3$ to 5, although there might be a
slight decrease in the fainter part. This suggested trend now becomes
clear thanks to the availability of deeper galaxy samples at all three
epochs: although the limiting absolute magnitude of our $z \sim 5$
sample is shallower than the lower-$z$ samples of
\citet{kdf2}, we see that at luminosities fainter than
$L^\ast_\mathrm{z=3}$ ($M_\mathrm{UV} \approx -21.0$) the number
density of LBGs steadily increases from $z=5$ to 3, while no number
density evolution is found at brighter magnitudes.

In order to be rigorous we need to quantify the strength and
statistical significance of this differential, luminosity-dependent
evolution.  However, we do not wish to compare the evolution of the
Schechter function parameters because the three Schechter parameters
($M^*$, $\phi^*$, and $\alpha$) are highly degenerate and so it is
difficult to physically interpret the meaning of their time-evolution.
Instead, it would be far better to study the evolution of the observed
number densities directly, but this is not simple 
since, at the three different epochs, the UVLF data span somewhat
different luminosity ranges, and because the magnitude bins are
centred at different $M_\mathrm{UV}$.  To overcome these problems, we
use the approach developed by \cite{kdf2} --- we investigate how far
the observed number densities at the three epochs deviate from a
fiducial model, for which we adopt the Schechter function fit to their
$z \sim3$ data.  In this way, for each data point, we construct
$\Phi(M,z) =\phi_{data}(M,z)/\phi_{fit}^{z\sim3}(M)$, which measure
the departure of the data from the model.  At $z \sim 3$, $\Phi(M)\sim
1$ at all $M$, reflecting the fact that the $z\sim 3$ Schechter
function is a good description of the $z\sim 3$ data. At other
redshifts, however, the $\Phi(M)$ deviate from unity, reflecting the
UVLF's evolution.  Next, at each epoch we compute $\bar{\Phi}(z)$ ---
i.e., the average of the $\Phi(M,z)$ --- for bins brighter (and
fainter) than $M_{UV} = M^*_{z=3} = -21.0$ mag.  Finally, the ratio of
the $\bar{\Phi}$ at two redshifts tells us the amount of number
density evolution that a given population (bright or faint) undergoes.

\begin{figure*}
\resizebox{12cm}{!}{
\includegraphics{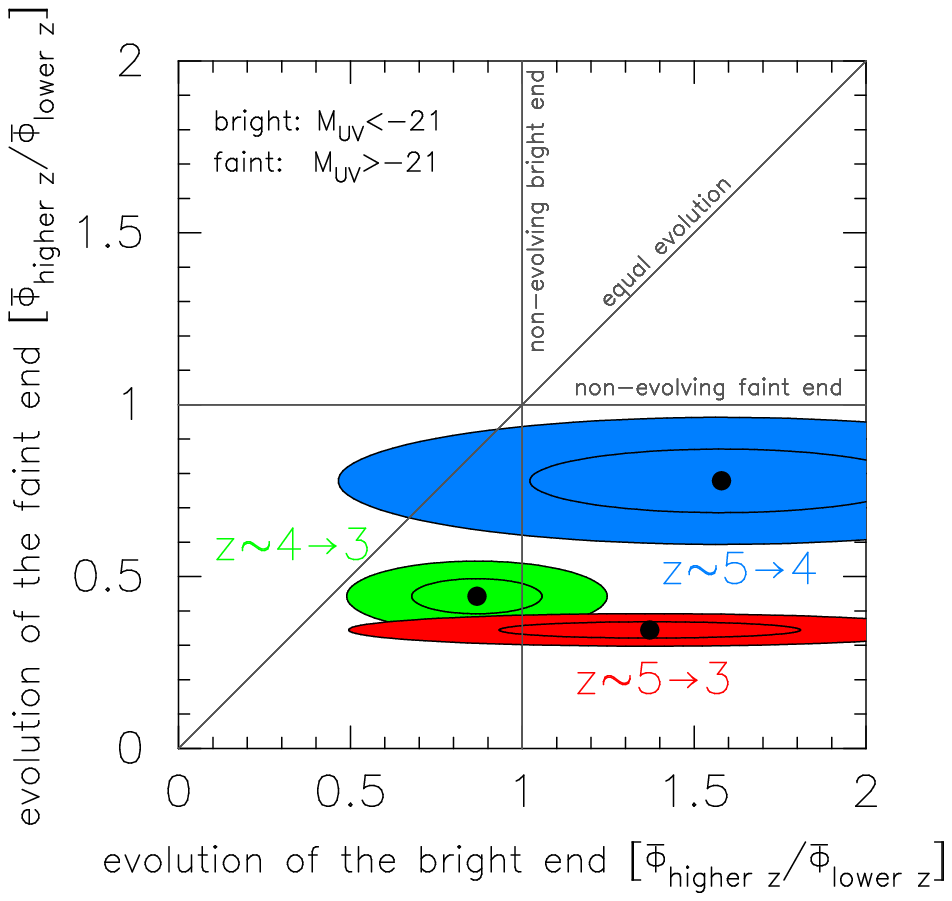}}
\caption{Luminosity-dependent evolution of the 
number density of galaxies.  The horizontal axis shows the change in
the number density of luminous galaxies ($M_{UV} < -21.0$ mag)
expressed as the ratio between the normalized average densities
$\bar{\Phi}$ at the redshift pairs indicated.  The vertical axis shows
the same quantity but for low luminosity galaxies ($M_{UV} > -21.0$
mag).  The error ellipses show 68\% and 95\% confidence regions, while
the straight lines represent three fiducial evolutionary scenarios as
labelled.  At $z \sim 5$ there are only $\sim 1/3$ as many faint
galaxies as at $z \sim 3$, and it is clear that the evolution of the
UVLF is luminosity-dependent over $z \sim 5 \rightarrow 3$ and $z \sim
4 \rightarrow 3$.  }
\label{fig:diffevsig}
\end{figure*}

The results of the above procedure are shown in
Fig.~\ref{fig:diffevsig} (see also Table~\ref{tbl:diffevsig}), which
shows the number density evolution of the bright end (horizontal axis)
and the faint end (vertical axis) of the UVLF.  The data show strong
evidence for evolution of the faint end of the UVLF: at $z \sim 5$
there are only 0.34 $\pm$ 0.02 as many faint galaxies as there are at
$z \sim 3$.  This represents a three-fold increase that is highly
statistically significant.  The evolution over $z \sim 5 \rightarrow
4$ is smaller (it is a 1/0.78 = 1.3-fold increase), but since the
cosmic time between $z=4.8$ and $z=4$ ($\sim$300 Myr) is half of the
interval between $z=4$ and $z=3$ ($\sim$600 Myr), the data are
completely consistent with the idea that the number of faint LBGs
increases at a constant rate from $z \sim 5$ to 3.  However, while the
faint end of the UVLF is evolving, there is no evidence for any change
in the number density of luminous ($M_\mathrm{UV}<-21.0$ mag) galaxies
over $z \sim 5 \rightarrow 4 \rightarrow 3$.  Indeed, the $z \sim
5\rightarrow 3$ data rule out a scenario --- indicated by the diagonal
line --- in which the number density of galaxies evolves uniformly at
all luminosities.  These $z\sim$5-based results are in excellent
agreement with the trends found at $z \sim 4 \rightarrow 3$ by
\citet{kdf2} (shown with green symbols), although our new $z \sim 5$
data both lengthen the cosmic time baseline and strengthen the
statistical significance of those earlier conclusions.

In summary, the comparison of our UVLF at $z \sim 5$ with those at
$z=3$--4 from \citet{kdf2} indicates that the evolution of LBG number
densities depends on UV luminosity.  Over the $\sim$1~Gyr from 
$z\sim 5 $ to $z \sim 3$, the number density of luminous LBGs
stays almost unchanged while there is a gradual and steady build-up in
the number density of faint ($M_\mathrm{UV} \ga -21$ mag) galaxies.

\subsection{Implications of UV Luminosity-dependent Evolution}

\subsubsection{Star Formation History of LBGs}

The UV luminosity-dependent evolution of UVLF has been already pointed
out by \citet{kdf2} for their LBG samples at $z \sim 3$ and 4. Our
data indicate that similar evolutionary trends were already in place
at earlier cosmic times.

\citet{kdf2} discuss three possible scenarios responsible for the
differential evolution of the UVLF between $z \sim 4$ and 3.  The
first one is the straightforward increase of the number of
low-luminosity LBGs at later epochs. They argue that, if the ages of
low-luminosity LBGs at $z \sim 3$ are mostly smaller than 600 Myr, the
increase in number density can be a natural consequence of the
emergence of large population of low-luminosity galaxies after $z \sim
4$. Sufficiently young ages of $z \sim 3$ LBGs have been suggested by
several authors. For example, \citet{shap01} used rest-frame UV to
optical SEDs of their (fairly bright) sample LBGs for comparisons with
template SEDs based on models with a simple constant star formation
history and derived a median age of 320 Myr, and results with similar
or younger age estimates have been also reported by other authors
(e.g., \citealt{sy98}; \citealt{pap01}; \citealt{iwt05}).  We should
note, however, that the ages estimated by SED fitting based on simple
star formation history models (constant or exponentially decaying SFH)
are strongly affected by recent star formation activity.  Thus the age
should be interpreted as time since the onset of most recent star
formation, and such age estimates may be incorrect if LBGs have
sporadic star formation histories.  As we have discussed in
\citealt{i03}, since the number density of LBGs in the luminous part
of the UVLF shows little change from $z \sim 5$ to 3, the small ages
of LBGs at $z \sim 3$ would imply that the star formation histories of
LBGs are sporadic, and that individual LBGs would have experienced
short bursts of star formation during their evolution.  Thus, if the
evolution of the fainter part of the UVLF represents the emergence of
a large population of low-luminosity LBGs at later epochs, it should
be interpreted as the increase in the number density of low-luminosity
population as a whole, and for individual galaxies a low-luminosity
LBG at $z \sim 4$ can be a luminous one at $z \sim 3$ and vice versa.

The change in the properties of starbursts, such as intensity, 
duration and duty cycles, can be the second possible cause of 
differential evolution.
It would be possible to explain the increase in the number
density of faint galaxies by introducing the differential change 
in the properties of starburst depending on UV luminosity 
(scenario B of \citealt{kdf2}).  
If intrinsically less luminous galaxies (unexplored
with current observations when in their low star formation state)
progressively spend longer time in the relatively luminous phase, they
can be observable by $z\sim3$, and as a consequence the number density
in the faint end of the UVLF could increase.  The luminous end would
not change if the starburst properties of luminous galaxies remain the
same.  Such scenario might be tested through the comparison of
starburst ages of LBG populations at different epochs.  Exploration of
the stellar populations of LBGs at $z \ga 4$ has become possible
recently thanks to the availability of rest-frame optical data
obtained through mid-infrared imaging with the Spitzer space
telescope, and several authors have reported the existence of massive
galaxies at $z \sim 6$ (e.g., \citealt{eyl05}; \citealt{mob05};
\citealt{yan06}).  Measurement of stellar masses would to some extent
resolve the ambiguity for the star formation history of LBGs present
in analyses based solely on the rest-frame UV wavelengths.  A
systematic study based on a uniform selection scheme across the
redshift range $z=3$--5 would be an interesting project to pursue in
order to address these issues.

One of the major obstacles for precise estimation of LBGs' ages would
be the notorious degeneracy between age and dust in their effects on
galaxy SEDs.  Indeed, \citet{kdf2} raised the change of dust
properties as the third possible cause of differential evolution.  If
the property of dust changes differentially depending on UV
luminosity, this would also alter the shape of the UV luminosity
function.  Unfortunately, we know quite little about the properties of
dust in high-z star-forming galaxies, and usually one attenuation law
(such as \citealt{cal00}) is assumed for all galaxies in a high-$z$
sample. Thus, it could be difficult to discriminate whether it is age
or dust that changes by studying the discrete SEDs obtained through
the broad-band imaging data.

\subsubsection{Biased Galaxy Evolution}

We recently found that Lyman $\alpha$ equivalent widths of $z \sim
5$--6 LBGs (and Lyman $\alpha$ emitters) tend to depend on UV
luminosity \citep{and06}: high luminosity LBGs show little or no Lyman
$\alpha$ emission, while there are objects with $M_\mathrm{UV} > -21$
mag which show strong emission lines.  Along with the result presented
in this paper that the number density of luminous LBGs at $z \sim 5$
is almost the same as at later epochs, the UV luminosity dependence of
Lyman $\alpha$ equivalent widths is another indication that UV
luminous galaxies have different properties from fainter ones.  As
discussed in \citet{and06}, there are several possible causes of the
luminosity dependence of the strength of Lyman $\alpha$ emission, one
possible explanation being that luminous LBGs are embedded in dense
reservoirs of neutral hydrogen gas.  The existence of such large
reservoirs of neutral gas could be the cause of the large star
formation rates we see and also could be the underlying cause of the
heavy absorption of Lyman $\alpha$ emission by ionized gas surrounding
the star forming regions within galaxies.  Another possibility is the
large amount of dust and/or metals in UV-luminous LBGs which have
experienced significant amounts of star formation until $z \sim
5$. The existence of strong silicon and carbon interstellar absorption
lines \citep{and04} in the luminous objects in our sample of $z \sim
5$ LBGs supports this idea, although sample size is still quite
limited.  A third indication of the UV luminosity dependence is the
relatively small number of bright objects with blue rest-frame UV
colours in our sample (section 3.6 and Fig.~\ref{fig:col_mag}).  The
dependence on UV luminosity of the clustering strength of $z=3$--5
LBGs has also been reported by several authors (e.g., \citealt{gd01};
\citealt{sdf6}; \citealt{kas06}).  It suggests UV luminous LBGs reside
in relatively massive dark matter haloes.

The existence of a number density of UV luminous LBGs at $z \sim 5$
that is comparable to those at $z=3$--4, as well as the
luminosity-dependent properties of LBGs especially at $z \sim 5$,
support the idea of biased galaxy formation.  The luminous galaxies,
presumably hosted by relatively massive dark matter haloes which are
also reservoirs of huge amounts of neutral gas, might start bursts of
star formation at early epochs (i.e., $z > 5$). On the other hand, the
number density of faint objects ($M_\mathrm{UV} > -21$ mag) gradually
increases along with the increase in the number density of dark matter
haloes.
The UV luminous objects, which have SFRs as large as $100
M_{\sun}$/yr$^{-1}$ at $z \sim 5$, may become fainter at later times
because the duration of such enormous bursts of star formation can be
on order of 100 Myr at most given that significantly longer bursts
would result in masses that exceed $10^{11} M_{\sun}$ by $z \sim 3$.
However, some fraction of objects that are relatively faint at $z \sim
5$ would be ignited during the period between $z \sim 5$ and 3, and
this would make it possible to keep the number density of UV-luminous
LBGs as a whole almost constant over the intervening $\sim$1 Gyr.
Such mixing of LBG populations that represent different generations
across cosmic time may by $z \sim 3$ extinguish the dependence of dust
/ metal or of gas mass on UV luminosity, and this mixing may explain
the apparent absence of UV luminosity dependence of Lyman $\alpha$
equivalent widths and strengths of interstellar metal absorption lines
at these lower redshifts\citep{shap03}.  This biased formation
hypothesis could be tested by further analyses of chemical
compositions and stellar populations for LBGs at different redshifts
and luminosity ranges. 
Although attempts to tackle these issues have 
been started (e.g, \citealt{and04}; \citealt{eyl05}; \citealt{lab07}), 
deep follow-up observations in both 
spectroscopy and multi-wavelength imaging that cover many 
sample galaxies are required.

\subsection{Comparison with UVLFs at $z \ga 6$}

Recently, there have been several attempts to determine the UVLF at $z
\sim 6$ through deep observations in $i$ and $z$ bands (e.g.,
\citealt{bou03}; \citealt{bun04}; \citealt{yan04}; \citealt{shim05}). 
\citet{bou06b} made a compilation of very deep observations with HST 
(GOODS-N, GOODS-S, the Hubble Ultra Deep Field (UDF) and UDF pararell
ACS fields) and derived the $z \sim 6$ UVLF at $1350\AA$ down to
$M_{1350} \sim -17.5$ mag. They argued that there is a strong evolution 
in the number density of luminous ($M < -21$ mag) LBGs from $z \sim 6$ 
to $z\sim 3$, while the number density of faint galaxies is 
comparable to or even larger than that at $z \sim 3$. 
They claim that it could be caused by the brightening of
$M^\ast$ along cosmic time.  Also, \citet{bou06a} utilized deep
HST/NICMOS observations to find LBGs at $z \approx 7$--8 and concluded
that there is a deficiency of luminous galaxies at $z \approx 7$--8
compared to $z \sim 6$, suggesting that the rapid evolution of the
number density of luminous galaxies has started at $z \ga 7$.  Their
arguments are opposite from the differential evolution of UVLFs we
found for the redshift range 3--5, and these two sets of results would
be inconsistent if one assumes that the number density evolution from
$z \sim 6$ to 3 proceeds continuously.  However, we should note that
the number densities in the fainter part of the $z \sim 6$ UVLF
reported by several authors have a fairly large scatter (see Table 13
and Figure 14 in \citealt{bou06b} for a summary), even though most of
these studies are based on similar data sets taken from deep HST
surveys.  Also, current deep HST based studies for $z \sim 6$ LBGs
suffer from severe limitations in survey volumes. The GOODS survey
area which \citet{bou06b} relied on to correct for the field-to-field
variations is just a fourth of the effective survey area presented
here. Thus the $z \sim 6$ UVLF and its Schechter parameters 
would be still need to be further examined. 
Although the number of objects with $z_{850} < 25.5$ mag (which
roughly corresponds to $M^\ast_\mathrm{z=3}$) detected in the GOODS
area by \citet{bou06b} (8 objects) is inconsistent with no evolution
from $z \sim 5$ even considering the dimming from $z \sim 5$ to 6 and
the difference in the rest-frame wavelength traced by the $z$-band
filter (which may lead to $\sim 0.1$ mag of dimming), 
it is still premature to draw a conclusion that whether the 
UVLF evolution from $z \sim 6$ to 3 is due to the brightening of 
$L^\ast$. Further studies of $z \sim 6$ LBGs, especially those based on 
wider survey area are indispensable (cf. \citealt{shim05}).

\section{UV Luminosity Density}

\subsection{UV Luminosity Density at $z \sim 5$}

Using the updated luminosity function of $z \sim 5$ LBGs we can
calculate the UV luminosity density at that epoch more reliably than
we did in \citet{i03} thanks to the deeper and wider data.  The
luminosity density at a target epoch due to light emitted from
galaxies brighter than $L_{lim}$ can be calculated by the integration
of the LF as

\[
 \rho(L>L_{lim}) = \int^{\infty}_{L_{lim}} L \phi(L) dL.
\]

We take two ranges of integration. One lower limit in the integration
is $M_\mathrm{UV} \leq -18.5$, which corresponds to $0.1
L^\ast_\mathrm{z=3}$, and the other is $L_{lim}=0$.  The formal lower
luminosity limit has been used by several previous studies (e.g.,
\citealt{steidel99}; \citealt{kdf3}; \citealt{bun06}), and 
corresponds to the limiting magnitude of the $z \sim 3$ LBG sample of
\citet{kdf2}.  At $z \sim 5$, we have to extrapolate our observed UVLF
more than 1 magnitude in order to obtain the UV luminosity density
integrated down to $M_\mathrm{UV} \leq -18.5$ mag.  With the
extrapolation of the integral down to zero luminosity ($L_{lim}=0$) we
want to estimate the total UV luminosity density emitted from
star-formation galaxies into inter-galactic space.  By comparing the
UV luminosity densities down to $0.1 L^\ast_\mathrm{z=3}$ with the
total ones we are able to estimate the contribution from faint
objects.  The steepness of the UVLF faint-end slope has a crucial
importance for the total UV luminosity.  As we have stated in section
4.2, the faint-end slope of the UVLFs is still not well constrained by
observations (especially for higher redshift, i.e., $z > 4$), and it
introduces a large uncertainty in the estimation of total UV
luminosity density.

\begin{figure*}
\resizebox{12cm}{!}{
\includegraphics{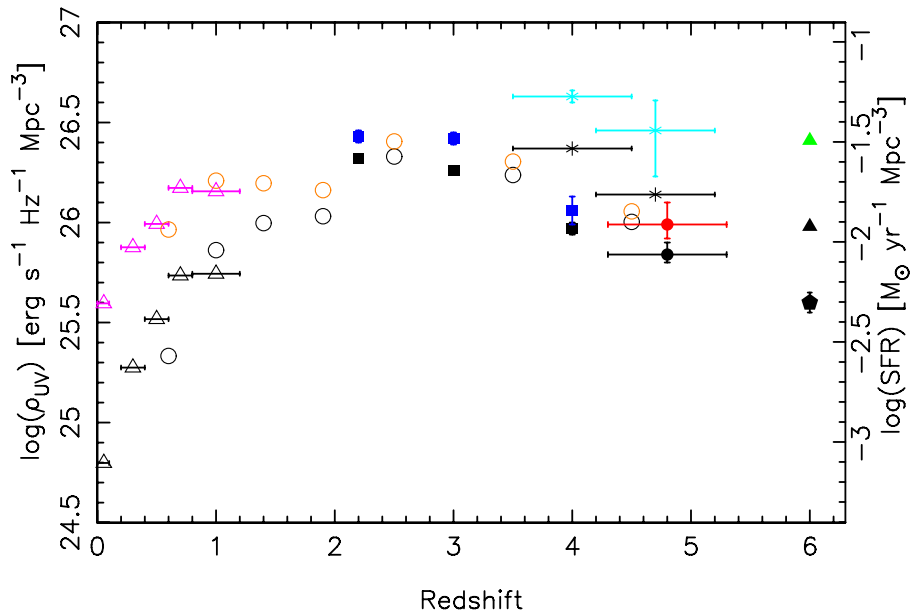}}
\caption{UV luminosity densities at various redshifts. 
Coloured symbols: UV luminosity densities integrated down to zero
luminosity.  Black-and-white symbols: UV luminosity densities
integrated down to $L_{\mathrm UV} = 0.1 L^\ast_{z=3}$ ($M_{\mathrm
UV} = -18.5$ mag).  Filled circles at $z=4.8$: this work, using the
UVLF derived from the combination of our two survey fields.  Open
triangles at $0 < z \leq 1$: based on UVLFs by Takeuchi et al. (2005).
Filled squares at $2.2 < z < 4$: Sawicki and Thompson (2006b).
Crosses at $z=4.0$ and $z=4.7$: Ouchi et al. (2004).  Filled triangles
at $z=6$: Bouwens et al. (2006).  Pentagon at $z=6$: Bunker et
al. (2006).  Open circles: Gabasch et al. (2004).  For published data
points error bars are shown only when authors provide error values for
UV luminosity density at these specific integration ranges
($L_{\mathrm UV} \ga 0.1 L^\ast_{z=3}$ and $L>0$).  The right vertical
axis indicates the comoving star formation density converted from UV
luminosity density.  The conversion factor in the logarithmic scale is
27.903 (Madau et al. 1998).}
\label{fig:uvdensity}
\end{figure*}

Fig.~\ref{fig:uvdensity} shows with filled circles the UV luminosity
densities obtained from our data at $z = 4.8$.  The integrations are
made using the best-fit Schechter function for the final UVLF
described in section 4.2, and error bars indicate $1\sigma$ errors
calculated using Schechter functions for perturbed data generated to
obtain 68\% confidence parameter ranges (see section 4.2).
Table~\ref{tbl:uvdensity} lists the UV luminosity densities at $z
\sim 5$ calculated using the final UVLF and the UVLF for the HDF-N
region.  Our survey only reaches to $M_\mathrm{UV} < -19.8$, and thus
68\% (49\%) of UV luminosity density integrated down to $M_\mathrm{UV}
\leq -18.5$ ($L_{lim}=0$) comes from this observed luminosity range.
Since our updated UVLF presented here is consistent with our previous
(\citet{i03}) one in the luminosity range covered by the previous
data, the UV luminosity density at $z \sim 5$ also shows little change
from the value in \citet{i03}.

\subsection{Evolution of UV Luminosity Density}

In Fig.~\ref{fig:uvdensity} we also show results for various redshifts
by other authors.  For $z \leq 1$, we show UV luminosity density data
from the combined sample of GALEX and VIMOS-VLT Deep survey
(\citealt{tak05}, who took the UVLF data from
\citealt{arn05}).  The UV luminosity densities at $z \sim 2$, 3
and 4 are from \citet{kdf3}. Their results are based on UVLFs in
\citet{kdf2}.  Redshift 4.0 and 4.7 results by \citet{sdf5} are
also shown.  At redshift 6 we show results by two different authors:
triangles are from \citet{bou06b} and the pentagon is from \cite{bun06}
(only the value for $L \geq 0.1 L^\ast_\mathrm{z=3}$ is provided in
the latter paper).  The rest-frame wavelengths of all of these
observations lie within 1350\AA--1700\AA.  For these published values,
we only show error bars when errors for these specific integration
ranges ($L\geq 0.1 L^\ast_\mathrm{z=3}$ and $L > 0$) are provided in
the papers, since we do not know the range of Schechter LF parameters
for $1\sigma$ confidence level which is required to precisely
calculate luminosity density error values.  No extinction correction
has been applied.

In the redshift range $z \ga 4$ the data points show fairly large
discrepancies. As has been pointed out in \citet{kdf3}, the value
obtained by the SDF for $z \sim 4$ is about 2.5 times larger than that
by the KDF for the integration down to $L=0.1 L^\ast_\mathrm{z=3}$ and
the disagreement reaches a factor of 3.7 when extrapolated to zero
luminosity. The situation is similar at $z \sim 5$ where SDF estimates
of UV luminosity density are larger than our results by factors of
2--3. The large luminosity density values by \citet{sdf5} come from
relatively large number densities in the fainter part of their UVLF
(see Fig.~9 of \citealt{kdf2}). By unconstrained Schechter function
fitting they obtained the best-fitted faint end slope of $\alpha =
-2.2$ for $z \sim 4$ LBGs, which is significantly steeper than the one
by \citet{kdf2} ($-1.26$) and the values estimated for $z \sim 3$ LBGs
($\approx -1.6$ to $-1.4$; \citealt{steidel99}; \citealt{kdf2}).  In
addition to UV luminosity density estimates based on this steep
$\alpha = -2.2$ slope, they also examined cases with a fixed UVLF
slope value of $-1.6$ to calculate UV luminosity densities at
$z=4$--5, and all results from \citet{sdf5} shown in
Fig.~\ref{fig:uvdensity} are based on calculations assuming the faint
end slope fixed to $-1.6$.

The \citet{sdf5} $z \sim 5$ number density at $M_\mathrm{UV}
\approx -20.8$ mag (the faintest data point for them) is comparable to
ours, but since their brighter part of the UVLF has smaller number
densities than does our LF, their best-fitted $M^\ast$ is fainter and
$\phi^\ast$ is larger than ours. Their assumed faint-end slope
($-1.6$) is also steeper than our best-fit value ($-1.48$).  As a
result, the contribution from fainter objects (extrapolated beyond the
observed limiting magnitude) to the UV luminosity becomes larger than
that in our data.  The large differences between the \citet{sdf5}
total UV luminosity densities and those integrated down to $0.1
L^\ast_\mathrm{z=3}$, at both $z \sim 4$ and $z \sim 4.7$, also
indicate that in their results the UV luminosity density comes mainly
from unexplored faint populations.  Because the bulk of UV luminosity
emitted into inter-galactic space is from galaxies with UV luminosity
$L_\mathrm{UV} \la L^\ast$, luminosity ranges probed by \citet{sdf5}
and \citet{i03} are too shallow to definitely measure the total UV
luminosity.  This is also the case for our result presented in this
paper, since our data are deeper but still only reach to $0.3
L^\ast_\mathrm{z=3}$.  For $z \sim 6$, the studies presented here both
(at least partly) use the common Hubble Ultra Deep Field (UDF) data
but, despite this, their results differ from each other.  To discuss
in detail what causes this discrepancy at $z \sim 6$ is beyond the
scope of this paper, but considering the facts that $z \sim 6$ LBGs
are primarily selected by one $i-z$ colour\footnote{Some part of UDF
has very deep near-infrared imaging with NICMOS \citep{bou06b}.}, and
that in both studies UV continuum luminosity estimated from $z'$-band
magnitudes may suffer attenuation by intergalactic neutral hydrogen,
this diversity at $z \sim 6$ may be caused by the uncertainty in
sample selection and/or correction for UV luminosity
(cf. \citealt{kdf3}).

Despite the uncertainty in the higher redshift ranges, it would be
meaningful to examine the overall change of UV luminosity density with
redshift using the currently available data in order to see the cosmic
evolution of star formation activity.  The trend suggested by
\citet{gab04b} based on their photometric redshift sample,
is that at $z \ga 3$ the UV luminosity density is slowly declining
with increasing redshift.  This trend is consistent with the
combination of \citet{kdf2} and our results, and the UV luminosity
density values of \citet{gab04b}, determined from the UVLF for
galaxies with photometric redshift $4<z<5$, also agree reasonably well
with ours results.  If we restrict ourselves to see only data points
with integration down to $0.1 L^\ast_\mathrm{z=3}$ (black and white
points in Fig.~\ref{fig:uvdensity}), which are less affected by the
uncertainty of the UVLF faint-end slope, then all the data in
Fig.~\ref{fig:uvdensity} show a consistent picture that the UV
luminosity density (i.e., comoving star formation density, if we
neglect the effect of dust attenuation) is low at $z \sim 6$ (when the
cosmic age is $\approx0.9$ Gyr) and gradually increases until $z \sim
2$--3 (when the cosmic age is 2--3 Gyr).  From that epoch on the UV
luminosity density declines rather rapidly until now.

This overall cosmic evolution of UV luminosity density might indicate
that some kind of transition in star-forming galaxies has happened at
$z \approx 2-3$.  At $z < 2$, it has been pointed out that in the
later epochs there is an absence of massive galaxies with large SFR
while less massive galaxies continue to form stars.  This trend is
called ``down-sizing'' and recent development of deep and wide imaging
surveys and intensive surveys using multi-object spectrographs (e.g.,
\citealt{kod04}; \citealt{treu05}; \citealt{jun05};
\citealt{per05}) made this trend clearer than when it was first 
suggested \citep{cow96}.  On the other hand, in this paper we have
argued that there is differential galaxy evolution at $z \ga 3$, i.e.,
that the galaxy population with high SFRs maintains its number density
from early times, $z \sim 5$, until $z \sim 3$, while the number
density of galaxies with smaller SFRs gradually increases.  Moreover,
we also discussed that this differential evolution could be an
indication of biased galaxy evolution in the early universe.  The
break in the UV luminosity density around $z \approx$2--3 may indicate
the transition from biased galaxy evolution to down-sizing.  In other
words, the formation of massive galaxies, i.e., the active assembly of
stellar mass within the massive dark matter haloes, might have started
at $z>6$ and continued at roughly constant rate for more than 2 Gyr,
with the bulk of stellar mass assembly in such massive haloes
terminating around $z \sim 2$.  Although the observational evidence of
down sizing has accumulated rapidly, the biased galaxy evolution at
higher redshift is still just a hypothesis.  Moreover, currently we
only deal with rest-frame UV properties which are dominated by the
effects of short time-scale starbursts. In order to investigate what
is happening at $z > 3$ and to relate that activity to down-sizing in
the lower redshift range, we need information on masses (either
stellar or of dark matter haloes).

\subsection{Contribution to the Ionizing Background Radiation}

UV luminosity density can be used to estimate the contribution of
stellar sources to the cosmic ionizing background radiation under an
assumption about the ratio of ionizing radiation flux escaping from
galaxies to non-ionizing UV radiation.  In \citet{i03} we argued that
if the escape fraction is as large as 0.5, then ionizing radiation
from star-forming galaxies and from QSOs at $z \sim 5$ may be
sufficient to keep the universe at that redshift ionized.  Several
studies of galaxies at $z \sim 6$ have discussed whether the UV
luminosity density of $z \sim 6$ galaxies is sufficient to keep the
universe ionized (e.g., \citealt{bou03}; \citealt{bou06b};
\citealt{yan04}; \citealt{bun06}).   However, since the escape 
fraction of ionizing photons from galaxies is still almost unknown
observationally (e.g, \citealt{steidel01}; \citealt{gia02};
\citealt{ino05}; \citealt{shap06}), such arguments cannot be
definitive.  Additionally, the uncertainty in the UVLF, especially in
its faint-end slope, makes the estimates of ionizing radiation
difficult.  On the other hand, \citet{ino06} used the evolution of the
UV luminosity density presented in Fig.~\ref{fig:uvdensity}, emission
from QSOs, IGM opacity model and ionizing background radiation in the
redshift range 0 to 6 to predict the escape fraction of star-forming
galaxies.  They found that evolution of the escape fraction --- i.e.,
higher ($\ga 0.2$) escape fraction at higher redshift ($z \geq 4$) ---
may be required.  Unfortunately, the direct observation of ionizing
radiation at $z \ga 4$ is extremely difficult due to the increasing
opacity of the IGM.  We should first investigate the properties of
ionizing radiation from galaxies at $z \la 3$ to know the typical
value of the escape fraction in the high-redshift universe and its
relation to other physical properties of galaxies.

\section{Summary and Conclusions}

In this paper we have reported the construction of a large sample of
Lyman break galaxies (LBGs) at $z \sim 5$ and their UV luminosity
function (LF), obtained from deep large-area Subaru observations in
two blank fields, namely, the Hubble Deep Field-North and the
J0053+1234 regions.  Our main findings are as follows:

\begin{itemize}
\item Over an effective survey area of 1290 arcmin$^2$,
we obtained 853 LBG candidates at $z \sim 5$ with $z' < 26.5$ mag.
Spectroscopic redshift data for galaxies at lower redshifts ($z < 4$)
from published redshift surveys in the target fields as well as the
results of follow-up spectroscopy for a subset of our LBG candidates
confirm that our colour selection criteria are defined so that
star-forming galaxies at $z \sim 5$ are efficiently selected and
contaminations from objects at lower redshift are properly eliminated.

\item In the bright end of the UVLF  
there is no significant change in the number density from $z \sim 5$
to 3, while we found a gradual increase from $z\sim5$ to 3 in the LF's
fainter part ($M_\mathrm{UV} > -21$ mag). Such differential evolution
of the UVLF from $z \sim 4$ to 3 has been pointed out by
\citet{kdf2}, and our present work shows that this trend was already
underway at earlier cosmic times.

\item We found a deficiency of galaxies with 
blue UV colours ($I_c-z' < 0$) at bright magnitudes ($z' \la 24.5$
mag). This trend is significant even if we take photometric errors
into account.

\item The UV luminosity density at $z \sim 5$,
calculated by integrating the UVLF, is found to be
$38.8^{+6.7}_{-4.1}$\% of that at $z \sim 3$, for the luminosity range
$L>0.1 L^\ast_\mathrm{z=3}$.  By combining our results with those from
the literature, we found that the cosmic UV luminosity density is
slowly increasing with cosmic age from $z \sim 6$, that it then marks
its peak at $z=3$--2, and then declines until the current epoch.

\end{itemize}

We have discussed the origins of the differential evolution of the
UVLF along cosmic time and suggested that our observational findings
are consistent with the biased galaxy evolution scenario. A galaxy
population hosted by massive dark haloes starts active star formation
preferentially at early times and maintains its activity until $z \sim
3$, while less massive galaxies increase their number density later.
This idea is supported by other observational findings, including, (1)
the deficiency of UV luminous objects with blue UV colours, (2) the
fact that UV luminous objects at $z \sim 5$ have smaller Lyman
$\alpha$ equivalent widths compared to fainter objects, and (3) the
dependence of clustering strengths on UV luminosity which suggests
that UV luminous objects are more likely to be hosted by massive dark
matter haloes.  Further investigations of the properties of LBGs,
especially at higher redshifts ($z > 4$) and with wider dynamic ranges
in luminosity and wavelength, are required to test the reality of the
biased galaxy evolution scenario.

\section*{Acknowledgments}

We thank staff members of the Subaru Telescope and 
the Subaru Mitaka Office 
for their support during observations, especially 
Drs. Y. Komiyama, A. Tajitsu and H. Furusawa who 
assisted us as support scientists.
I.I. was supported by a Research Fellowship of the Japan 
Society for the Promotion of Science (JSPS) for Young Scientists 
during some part of this research, and he is supported by a 
Grant-in-Aid for Young Scientists (B) from 
the Ministry of Education, Culture, Sports, Science 
and Technology of Japan (18740114). 
K.O.'s activity is supported by a Grant-in-Aid for scientific 
research from JSPS (17640216). 
M.S. and K.O. acknowledge the program of Invitation Fellowship 
for Research in Japan by JSPS. 
Part of the data reduction was carried out on the ``sb'' computer 
system operated by the Astronomical Data Analysis Center (ADAC) and 
Subaru Telescope of the National Astronomical Observatory of Japan. 
I.I. thanks Okayama Information Highway and the National Institute of 
Information and Communications Technology (Japan Gigabit Network)
for their support on high-speed network connection for data transfer
and analysis in the Okayama Astrophysical Observatory.
We would like to express our acknowledgment to the indigenous 
Hawaiian community for their understanding of the significant 
role of the summit of Mauna Kea in astronomical research.



\begin{table*}
\centering
\begin{minipage}{140mm}
\caption{A summary of observations for HDF-N.
Limiting magnitude is for 5 $\sigma$ with 1.2\arcsec diameter aperture.}
\label{tbl:hdfnobs}
\begin{tabular}{@{}lccccc}
\hline
Filter & Single Exposure Time & Number of shots & Total integration time & Limiting mag. &  Seeing \\
& sec &  & sec & & \arcsec \\
\hline
$V$   & 720--1,440 & 14 & 24,720 & 28.17 & 1.11\\
      &        600 & 10 &        &       &     \\
$I_c$ & 180--300   & 11 &  6,180 & 26.86 & 1.13\\
      & 360--420   & 10 &        &       & \\
$z'$  & 120--215   & 21 & 11,935 & 26.55 & 1.13\\
      & 180--280   & 41 &        &       & \\
\hline
\end{tabular}
\end{minipage}
\end{table*}

\begin{table*}
\centering
\begin{minipage}{140mm}
\caption{A summary of observations for J0053+1234.
Limiting magnitude is for 5 $\sigma$ with 1.2\arcsec diameter aperture.}
\label{tbl:j0053obs}
\begin{tabular}{@{}lccccc}
\hline
Filter & Single Exposure Time & Number of shots & Total integration time & Limiting mag. &  Seeing \\
& sec &  & sec & & \arcsec \\
\hline
$V$   & 600 & 23 & 13,800 & 27.79 & 0.89\\
$I_c$ & 200 & 10 &  2,000 & 26.39 & 0.87\\
$z'$  & 240 & 28 &  6,720 & 26.24 & 0.88\\
\hline
\end{tabular}
\end{minipage}
\end{table*}

\begin{table*}
\centering
\begin{minipage}{140mm}
\caption{Cross identifications of objects with previously 
published spectroscopic redshifts at $4.5<z<5.5$ in the 
HDF-N region.
}
\label{tbl:crossid_hdfn}
\begin{tabular}{@{}lllrrrrl}
\hline
ID & R.A.(J2000)$^a$ & Dec.(J2000)$^a$ & $z'$ mag. & $V-I_c$ & $I_c-z'$ & Redshift & Ref.$^b$ \\
\hline
F36219-1516  & 12:36:21.88 & +62:15:17.0 & 25.39 & 0.67    & 0.09  & 4.890 & 1\\ 
F36279-1750  & 12:36:27.74 & +62:17:47.8 & 26.05 & $>$2.32 &$-0.11$& 4.938 & 1\\
F36376-1453  & 12:36:37.62 & +62:14:53.8 & 21.97 & 3.25    & 0.78  & 4.886 & 1\\ 
HDF 4-439.0  & 12:36:43.84 & +62:12:41.7 & 25.29 & 1.48    &$-0.19$& 4.54  & 2\\ 
HDF 4-625.0  & 12:36:44.65 & +62:11:50.7 & 25.24 & $>$2.87 & 0.03  & 4.580 & 3\\
GOODS J123647.96+620941.7 & 12:36:47.93 & +62:09:41.7 & 24.32 & $>$3.31 & $-0.10$ & 5.186 & 4\\
ES 1         & 12:36:49.23 & +62:15:38.8 & 25.48 & $>$2.96 &$-0.16$& 5.190 & 5\\
HDF 3-0951.0 & 12:37:00.23 & +62:12:19.8 & 24.69 & 1.82    & 0.22  & 5.34  & 6\\
095819(A04-4)& 12:37:05.68 & +62:07:43.3 & 24.50 & 2.26    & 0.15  & 4.650 & 7\\
GOODS J123721.03+621502.1 & 12:37:21.00 & +62:15:02.1 & 23.32 & 2.36 & 0.56 & 4.761 & 8\\ 
144200(A04-2)& 12:37:57.49 & +62:17:19.0 & 24.08 & 2.40    & 0.07  & 4.695 & 7\\
129178(A04-7)& 12:38:04.36 & +62:14:19.7 & 24.29 & $>$3.28 & 0.21  & 5.183 & 7\\
104268(A04-1)& 12:38:11.32 & +62:09:19.4 & 24.02 & 2.22    & 0.11  & 4.517 & 7\\
148198(A04-8)& 12:38:16.63 & +62:18:05.3 & 24.50 & $>$3.19 & 0.42  & 4.615 & 7\\
149472(A04-6)& 12:38:25.52 & +62:18:19.7 & 24.87 & $>$3.02 & 0.14  & 4.857 & 7\\
139294(A04-5)& 12:38:28.96 & +62:16:18.8 & 24.39 & 1.96    & 0.20  & 4.667 & 7\\
\hline
\end{tabular}
$^a$
Positions are based on our astrometry with Suprime-Cam images, 
using USNO-B1 catalog (see section 3.2).\\
$^b$ 
1: \citet{daw01}, 
2: \citet{strn99}, 
3: \citet{fsoto99}, 
4: \citet{bar02}, 
5: \citet{daw02}, 
6: \citet{spn98}, 
7: \citet{and04}, 
8: \citet{cow04}.
\end{minipage}
\end{table*}

\begin{table*}
\centering
\begin{minipage}{140mm}
\caption{Cross identifications of objects with previously 
published spectroscopic redshifts at $4.5<z<5.5$ in the 
J0053+1234 region.}
\label{tbl:crossid_j0053}
\begin{tabular}{@{}lllrrrrl}
\hline
ID & R.A.(J2000)$^a$ & Dec.(J2000)$^a$ & $z'$ mag. & $V-I_c$ & $I_c-z'$ & Redshift & Ref.$^b$ \\
\hline
106426(A07-1) & 00:52:21.34 & +12:32:35.3 & 24.07 &  2.59 & 0.23 & 4.797 & 1\\
093014(A07-4) & 00:52:37.37 & +12:29:58.7 & 25.26 &$>$1.85& 0.16 & 4.491 & 1\\
091813(A07-3) & 00:52:39.88 & +12:29:44.1 & 25.03 &  1.98 & 0.06 & 4.391 & 1\\
104115(A07-2) & 00:52:43.27 & +12:32:08.2 & 24.22 &  1.75 & 0.16 & 4.267 & 1\\
CDFa G01      & 00:53:33.21 & +12:32:07.3 & 23.54 & 2.88 & 0.03 & 4.815 & 2\\
CDFa GD07     & 00:53:35.57 & +12:31:44.1 & 23.72 & 1.67 & 0.17 & 4.605 & 2\\
CDFb G05      & 00:53:51.28 & +12:24:21.3 & 24.31 & 2.70 & 0.29 & 4.486 & 2\\
\hline
\end{tabular}

$^a$
Positions are based on our astrometry with Suprime-Cam images, 
using USNO-B1 catalog (see section 3.2).\\
$^b$ 
1: \citet{and07}, 
2: \citet{steidel99}.
\end{minipage}
\end{table*}

\begin{table*}
\centering
\begin{minipage}{140mm}
\caption{Number of LBG candidates and UV luminosity function of 
LBGs at $z \sim 5$ for HDF-N.}
\label{tbl:hdfuvlf}
\begin{tabular}{@{}llrlrl}
\hline
\multicolumn{1}{c}{$z'$ mag} & abs. mag & Num & $V_{\mathrm{eff}}$&
 $\mathrm{N_{int}}$& \multicolumn{1}{c}{$\phi$(m)}\\
  & & & Mpc$^{-3}$ & & N/Mpc$^{-3}$/mag \\
\hline
23.0 -- 23.5 & $-$23.082 &   2 & $3.07\times 10^3$ &  0.79 &$1.55 \pm 0.85 \times 10^{-6}$ \\ 
23.5 -- 24.0 & $-$22.582 &   5 & $2.86\times 10^3$ &  2.79 &$3.04 \pm 0.91 \times 10^{-6}$ \\ 
24.0 -- 24.5 & $-$22.082 &  35 & $2.68\times 10^3$ &  6.06 &$4.25 \pm 0.66 \times 10^{-5}$ \\ 
24.5 -- 25.0 & $-$21.582 &  72 & $2.48\times 10^3$ & 12.58 &$9.42 \pm 1.05 \times 10^{-5}$ \\ 
25.0 -- 25.5 & $-$21.082 & 126 & $2.24\times 10^3$ & 24.94 &$1.78 \pm 0.15 \times 10^{-4}$ \\ 
25.5 -- 26.0 & $-$20.582 & 186 & $1.90\times 10^3$ & 24.05 &$3.36 \pm 0.25 \times 10^{-4}$ \\ 
26.0 -- 26.5 & $-$20.082 & 191 & $1.51\times 10^3$ & 20.33 &$4.43 \pm 0.33 \times 10^{-4}$ \\ 
\hline
\end{tabular}
\end{minipage}
\end{table*}

\begin{table*}
\centering
\begin{minipage}{140mm}
\caption{Number of LBG candidates and UV luminosity function of 
LBGs at $z \sim 5$ for J0053+1234.}
\label{tbl:j0053uvlf}
\begin{tabular}{@{}llrlrl}
\hline
\multicolumn{1}{c}{$z'$ mag} & abs. mag & Num & $V_{\mathrm{eff}}$ &
 $\mathrm{N_{int}}$& \multicolumn{1}{c}{$\phi$(m)}\\
  & & & Mpc$^{-3}$ & & N/Mpc$^{-3}$/mag \\
\hline
23.0 -- 23.5 & $-$23.082 &   3 & $2.51\times 10^3$ &  1.40 & $1.63 \pm 0.69 \times 10^{-6}$ \\
23.5 -- 24.0 & $-$22.582 &  10 & $2.38\times 10^3$ &  1.08 & $9.61 \pm 2.88 \times 10^{-5}$ \\
24.0 -- 24.5 & $-$22.082 &  22 & $2.28\times 10^3$ &  2.53 & $2.19 \pm 0.44 \times 10^{-5}$ \\
24.5 -- 25.0 & $-$21.582 &  79 & $1.99\times 10^3$ &  7.82 & $9.16 \pm 1.02 \times 10^{-5}$ \\
25.0 -- 25.5 & $-$21.082 & 122 & $1.68\times 10^3$ & 14.85 & $1.63 \pm 0.15 \times 10^{-4}$ \\
\hline
\end{tabular}
\end{minipage}
\end{table*}

\begin{table*}
\centering
\begin{minipage}{140mm}
\caption{UV luminosity function of 
LBGs at $z \sim 5$ obtained as a weighted average of results for two
survey fields. The number densities at $z'>25.5$ come only from 
the HDF-N region.}
\label{tbl:uvlfmean}
\begin{tabular}{@{}llr}
\hline
\multicolumn{1}{c}{$z'$ mag} & abs. mag & \multicolumn{1}{c}{$\phi$(m)} \\
  & & N/Mpc$^{-3}$/mag \\
\hline
23.0 -- 23.5 & $-$23.082 & 1.60$\pm0.75\times10^{-6}$\\
23.5 -- 24.0 & $-$22.582 & 7.02$\pm2.11\times10^{-6}$\\
24.0 -- 24.5 & $-$22.082 & 3.00$\pm0.53\times10^{-5}$\\
24.5 -- 25.0 & $-$21.582 & 9.26$\pm1.03\times10^{-5}$\\
25.0 -- 25.5 & $-$21.082 & 1.69$\pm0.15\times10^{-4}$\\
25.5 -- 26.0 & $-$20.582 & 3.36$\pm0.25\times10^{-4}$\\
26.0 -- 26.5 & $-$20.082 & 4.43$\pm0.33\times10^{-4}$\\
\hline
\end{tabular}
\end{minipage}
\end{table*}

\begin{table*}
\centering
\begin{minipage}{140mm}
\caption{Schechter function parameters of UVLF at $z \sim 5$, 
obtained with the data from the HDF-N region only and 
from both the HDF-N and J0053+1234 regions.}
\label{tbl:uvlfpar}
\begin{tabular}{@{}rlll}
\hline
Region & \multicolumn{1}{c}{$M^\ast$} & \multicolumn{1}{c}{$\phi^\ast$} & \multicolumn{1}{c}{$\alpha$} \\
\hline
               HDF-N & $-20.82^{+0.22}_{-0.22}$ & $7.7^{+1.9}_{-2.1}\times10^{-4}$ & $-1.05^{+0.35}_{-0.32}$ \\
HDF-N and J0053+1234 & $-21.28^{+0.38}_{-0.38}$ & $4.1^{+2.9}_{-3.0}\times10^{-4}$ & $-1.48^{+0.38}_{-0.32}$ \\
\hline
\end{tabular}
\end{minipage}
\end{table*}

\begin{table*}
\centering
\begin{minipage}{140mm}
\caption{Evolution of the number density for galaxies brighter and fainter than $M_{UV}= -21.0$.
}
\label{tbl:diffevsig}
\begin{tabular}{@{}lcc}
\hline
\multicolumn{1}{c}{Epochs} & bright end ($M_{UV}< -21.0$) & faint end  ($M_{UV}> -21.0$)\\
 & $\bar{\Phi}(higher z) / \bar{\Phi}(lower z) $&  $\bar{\Phi}(higher z) / \bar{\Phi}(lower z) $\\
\hline
$z \sim 5 \rightarrow 3 $ & 1.37 $\pm$ 0.44 & 0.34 $\pm$ 0.02 \\
$z \sim 5 \rightarrow 4 $ & 1.58 $\pm$ 0.56 & 0.78 $\pm$ 0.09 \\
$z \sim 4 \rightarrow 3 $ & 0.87 $\pm$ 0.19 & 0.44 $\pm$ 0.05 \\
\hline
\end{tabular}
\end{minipage}
\end{table*}

\begin{table*}
\centering
\begin{minipage}{140mm}
\caption{UV luminosity density calculated for the 
luminosity range $L_{\mathrm UV} \ga 0.1 L^\ast_{z=3}$ 
($M_{\mathrm UV} \leq -18.5$) and $L_{\mathrm UV} > 0$.}
\label{tbl:uvdensity}
\begin{tabular}{@{}clll}
\hline
\multicolumn{1}{c}{redshift} & $L_{\mathrm UV} \ga 0.1 L^\ast_{z=3}$ & $L_{\mathrm UV} > 0$ & Reference \\
\hline
$5.0^{+0.5}_{-0.5}$ & $25.81^{+0.04}_{-0.04}$ & $25.86^{+0.09}_{-0.06}$ & HDFN \\
$5.0^{+0.5}_{-0.5}$ & $25.84^{+0.06}_{-0.04}$ & $25.99^{+0.23}_{-0.09}$ & HDFN and J0053+1234\\
\hline
\end{tabular}
\end{minipage}
\end{table*}

\label{lastpage}

\end{document}